\documentclass[%
 reprint,
 amsmath,amssymb,
 aps,
]{revtex4-2}

\usepackage{graphicx}
\usepackage{dcolumn}
\usepackage{bm}
\usepackage{lipsum}
\usepackage{color,soul}
\usepackage{dsfont}
\usepackage[colorinlistoftodos]{todonotes}
\usepackage{booktabs}
\usepackage{makecell}
\usepackage{graphicx}
\usepackage{subcaption}
\usepackage{rotating}
\usepackage{adjustbox}

\begin{document}

\preprint{APS}

\title{Quadrupole-Octupole Residual Interaction \\ in the Proxy-$SU(3)$ Scheme }

\author{A. Restrepo}
 \altaffiliation{alejandro.restrepo19@udea.edu.co}
\author{J. P. Valencia}%
 \email{patricio.valencia@udea.edu.co}
\affiliation{Instituto de Física, Universidad de Antioquia, Calle 70 No. 52-21, Medellín, Colombia}%




\date{\today}

\begin{abstract}

The octupole deformation of atomic nuclei is a relevant research area given its implications in the nuclear structure and fundamental physics, however, inclusion of octupole degrees of freedom in the nuclear interaction has been explored little in $SU(3)$ symmetry-based models. In this article we expand the octupole operator $\mathcal{O}^3_\mu$ in the $SU(3)$ second quantization formalism and use it to formulate an octupole-octupole residual interaction $\mathcal{O}^3\cdot \mathcal{O}^3$ which is incorporated in Elliott's model Hamiltonian. We compute the matrix elements of this extended Hamiltonian and use them to calculate energy  levels and analyse the $B(E3; 3^-_1\rightarrow 0^+_1)$ transition strength of the isotope $^{224}$Th using the semi-microscopic algebraic quartet model (SAQM) based on the proxy-$SU(3)$ scheme.

\end{abstract}

\maketitle

\section{\label{sec:theory}INTRODUCTION}

The atomic nucleus is a primary component of ordinary matter and its structure is one of the main subjects of study in quantum mechanics. Since the nucleus is an extended body, it is a legitimate area of research to investigate its geometrical shape \cite{Heyde_2016, RSMackintosh_1977, nilsson1995shapes}, the interactions influencing it \cite{PhysRevC.64.037301} and its variation across isotopes as well as between excited states \cite{RevModPhys.83.1467, Martinou2021}. It is known that some nuclei organize in spherical shapes while others lower its energy through deformation. This phenomenon is described theoretically using the multipole moments where the most relevant for the atomic nucleus are the first four, \textit{viz.} dipole ($Q_1$), quadrupole ($Q_2$), octupole ($Q_3$) and hexadecapole ($Q_4$) moments. 

Over the years, several non-zero values of electric quadrupole moments have been reported \cite{STONE20161} which indicate axially symmetric ellipsoidal deformations of the nucleus. Negative values correspond to oblate deformations while positive ones to prolate deformations, with a clear prolate dominance observed across all nuclei \cite{PhysRevC.95.064326, PhysRevC.64.037301}. The largest quadrupole values are generally found in half-filled shells defined by the magic numbers $2, 8, 20, 28, 50, 82,$ and $126$. This breaking of spherical symmetry allows for rotational degrees of freedom to populate excited states observed as characteristic energy level patterns called rotational bands. 

For certain configurations of proton ($Z$) and neutron ($N$) numbers around $\approx 34, 56, 88$, and $134$ \cite{Butler_2016}, the nucleus lowers its energy further through a static octupole deformation \cite{annurev, RevModPhys.68.349}. This combined with a prolate quadrupole moment gives the nucleus a so-called “pear shape"\cite{butlerpear, pancholi2020pear}. Such mirror asymmetric shape causes a displacement between the centers of charge and mass, creating an electric dipole moment as well. Although few direct measurements of the $Q_3$ moments have been reported in the literature, certain low energy spectra, enhanced $E3$ and $E1$ transition probabilities serve as indirect indicators of octupole deformation \cite{Butler_2016}. 

\begin{figure*}[t]
    \centering
    \begin{minipage}[b]{0.49\textwidth}
        \centering
        \includegraphics[trim={3.9cm 4.5cm 4cm 0}, clip, width=1\textwidth]{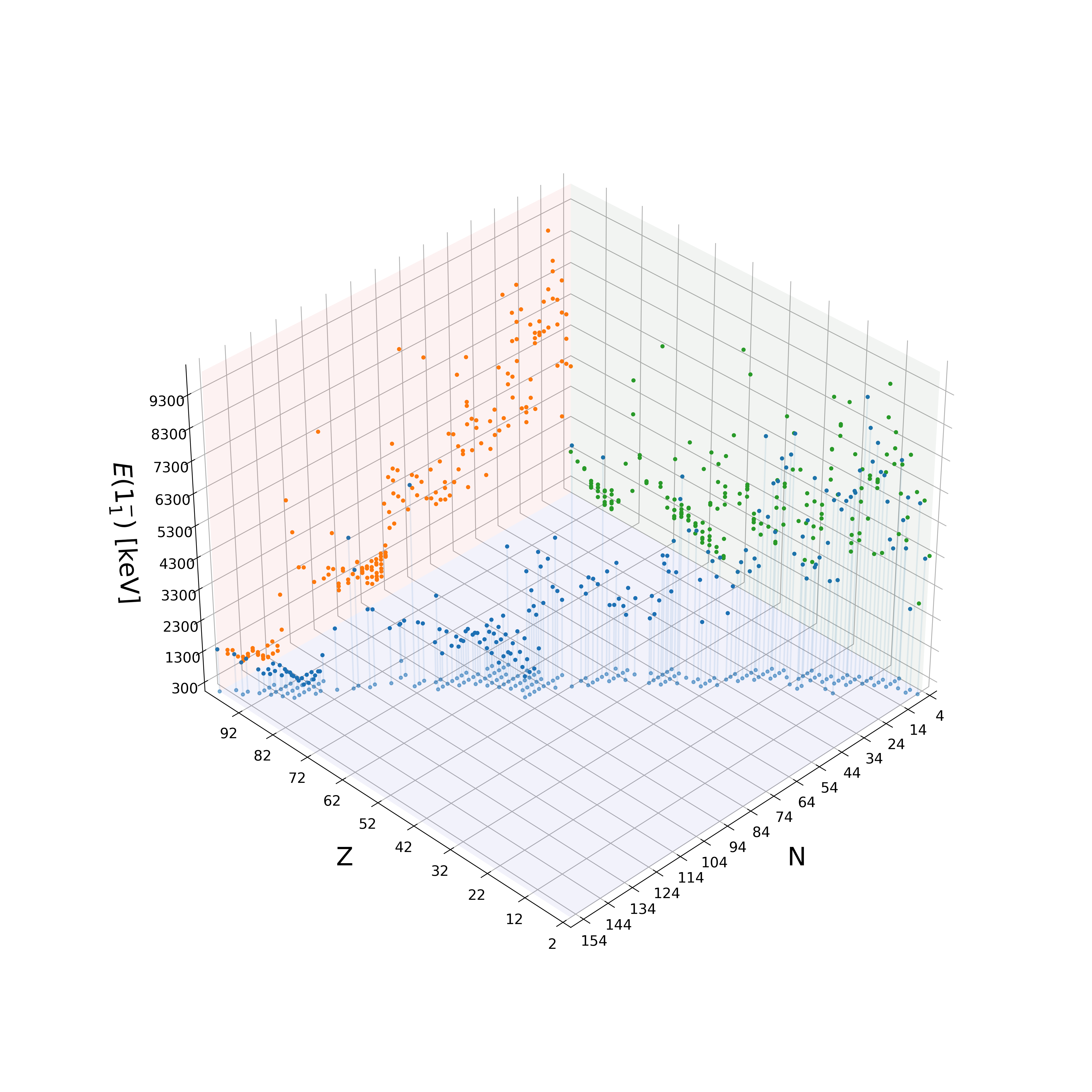}
    \end{minipage}
    \begin{minipage}[b]{0.49\textwidth}
        \centering
        \includegraphics[trim={3.9cm 4.5cm 4.2cm 0}, clip, width=1\textwidth]{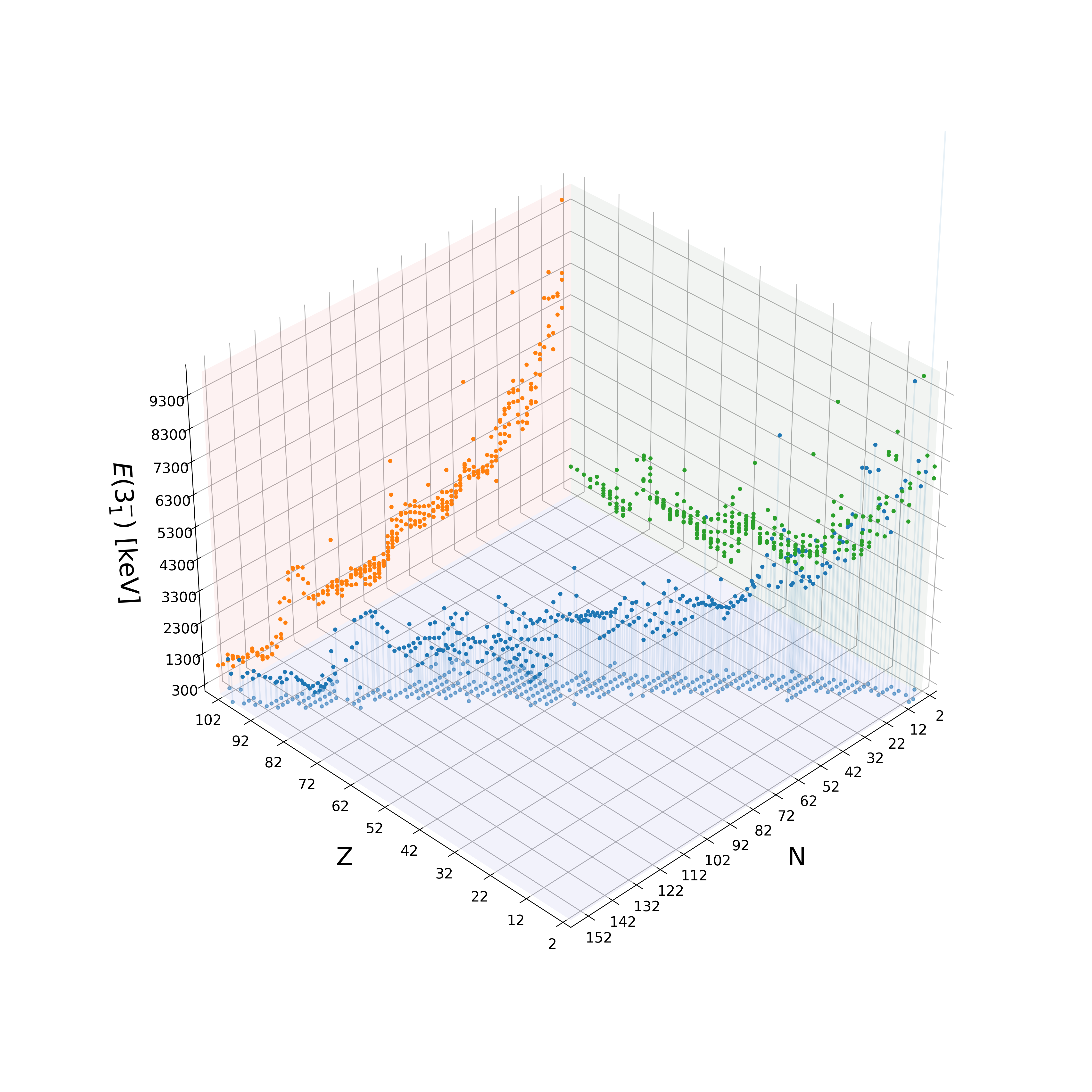}
    \end{minipage}
    \caption{Lowest $1^-$ (left) and $3^-$ (right) energy levels in even-even nuclei reported in ENSDF \cite{nndc}.}
    \label{E1E3}
\end{figure*}

\begin{figure}[p]
\includegraphics[trim={2cm 0 3cm 0cm}, clip, width=0.5\textwidth]{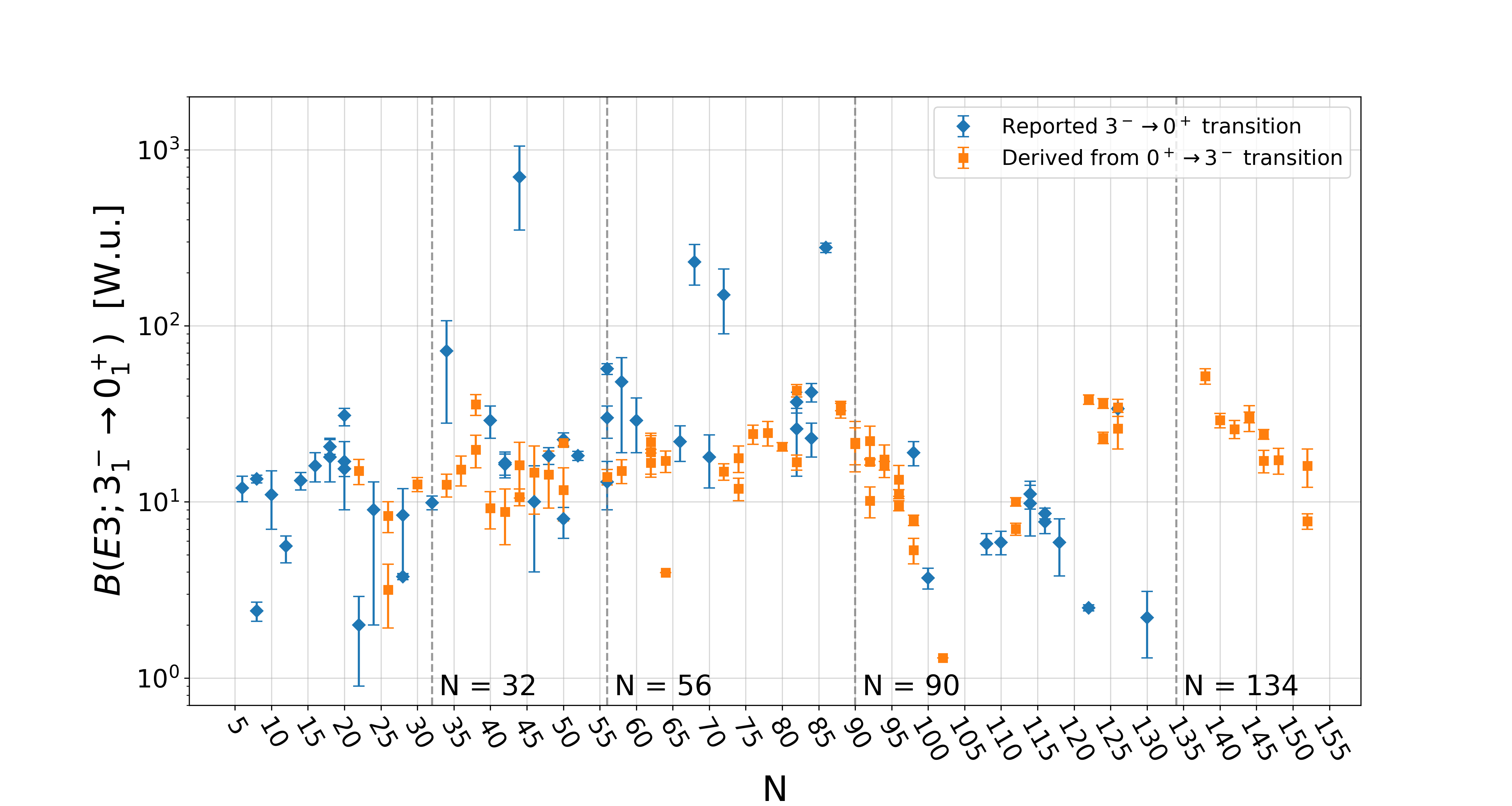}
\includegraphics[trim={2cm 0 3cm 2cm}, clip, width=0.5\textwidth]{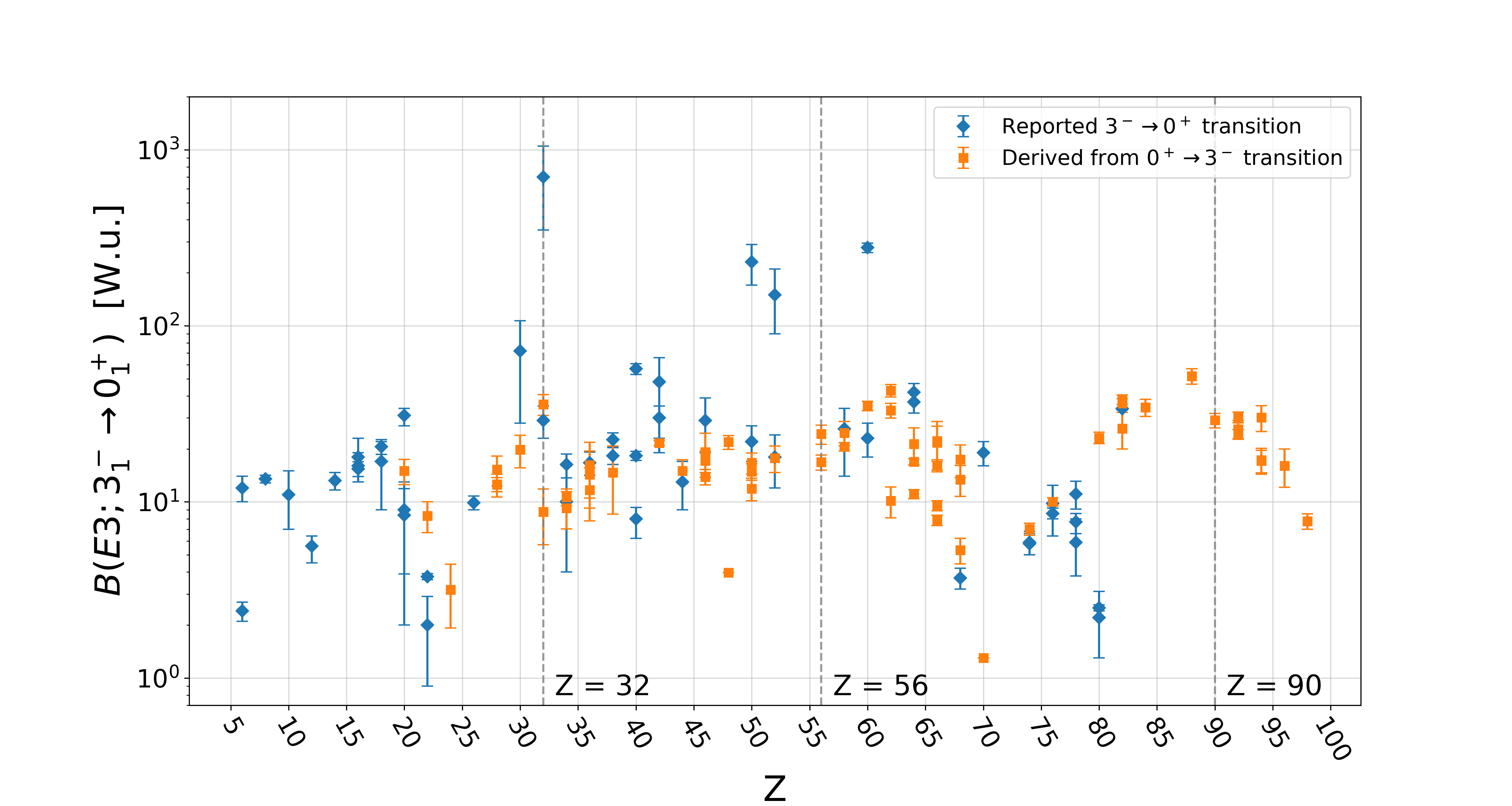}
\caption{ Evaluated $B(E3)$ values in even-even nuclei reported in ENSDF \cite{nndc}. The vertical lines indicate the octupole magic numbers obtained in \cite{Martinou_2024}. } 
\label{BE3N}
\end{figure}

In particular, for even-even nuclei, the presence of a negative parity band $L^\Pi = 1^-, 3^-, 5^-, ...$ near the ground state band $L^\Pi = 0^+, 2^+, 4^+, ...$ is a characteristic spectrum of “pear shape" deformation along with enhanced $3_1^- \rightarrow 0_1^+$ and $1_1^- \rightarrow 0_1^+$ transitions. In figure \ref{E1E3} are shown the energies of the lowest $1_1^-$ and $3_1^-$ excited states for even-even nuclei. It can be observed that the lowest values are found in the actinide region for the isotopes $^{224}$Ra, $^{224}$Th, $^{226}$Th, $^{222}$Ra, $^{226}$Ra and $^{228}$Th in increasing order up to 0.4MeV. In figure \ref{BE3N} are shown the evaluated $B(E3;3_1^- \rightarrow 0_1^+)$ values where a maximum can be found around that region. Previous calculations \cite{PhysRevC.89.024312} with relativistic Hartree–Bogoliubov (RHB) theory  predicted stable octupole deformations for isotopes $^{222-226}$Ra and $^{224-228}$Th suggesting them as the best even-even candidates for research on “pear shape" deformation. 

One of the main interests of studying octupole collectivity in nuclei goes further than the nuclear structure itself, having implications in the strong nuclear interaction and the symmetries of nature \cite{POSPELOV2005119, PhysRevLett.102.101601}. Experimental efforts are being made to measure $Q_1$, $Q_3$ and $E3$ transitions that will provide more insight about fundamental physics \cite{Gaffney2013-df, TFThorsteinsen_1990}. In this article we continue the work on the isotope $^{224}$Th in the semi-microscopic algebraic quartet model (SAQM) \cite{CSEH2015213} based on the proxy-$SU(3)$ scheme \cite{PhysRevC.95.064325, sym15010169} started in \cite{PhysRevC.101.054306, restrepo2024} by including a residual octupole-octupole interaction in Elliott's Hamiltonian. An analysis of octupole effective charges and $B(E3;3_1^- \rightarrow 0_1^+)$ transition are made as well. We begin explaining the two-shell model space of the system required for the residual octupole interaction to be defined, then show the expansion of the octupole operator in terms of $SU(3)$ irreducible tensors and the calculations for the low-energy spectrum of $^{224}$Th along comparison with observed energy levels.

\section{MODEL SPACE}

Since the octupole operator $r^3Y^3_{\mu}(\Omega)$ has negative parity, at least two consecutive shells are required for its minimal realization \cite{Isacker_2016}. We will focus on heavy even-even nuclei in the actinide region so that proton ($\pi$) and neutron ($\nu$) degrees of freedom must be distinguished. These two statements imply that the full model space has to be composed of two-shells which represent the valence and excited shells states for each proton and neutron spaces ($\eta_\pi = 5, 6$ and $\eta_\nu = 6, 7$ respectively) along with their couplings to total irreducible representations (irreps). 

The spectrum generating algebra (SGA) \cite{frank2008symmetries} will be $U(2(\Omega_{\sigma-}+\Omega_{\sigma+}))$ for each proton and neutron spaces where $\sigma = \pi, \nu$, the sub-indexes $-$ and $+$ indicate lower and upper shells respectively and $\Omega = (\eta+1)(\eta+2)/2$. Since nucleons are fermions, their SGA irreps must be totally anti-symmetric given by $\{1^{N_\sigma}\}$ where $N_\sigma$ represents the number of particles. This algebra can be separated into spin and spatial parts as a product $\Big\{U(\Omega_{\sigma-})\otimes U(\Omega_{\sigma+})\Big\}\otimes\Big\{ SU_{\sigma-}(2)\otimes SU_{\sigma+}(2)\Big\}$. The corresponding spatial representations will be denoted as $[f]_{\sigma_-}\otimes [f]_{\sigma_+}$.
In the SAQM, the ground state ($0\hbar\omega$) will be given by setting the upper shell empty in a configuration like $[2^I]_{\sigma-}\otimes[0]_{\sigma+}$ where $I$ denotes the number of quartets \cite{CSEH2015213, PhysRevC.101.054306}. Excited ($1\hbar\omega$) states will be given by occupying one orbital in the excited shell in the proton or neutron space (not simultaneously) denoted as $[2^{I-1},1]_{\sigma-}\otimes[0]_{\sigma+}$. These irreps guarantee that the quartet symmetry $[4^I]$ for the total proton-neutron space is maintained. The $U(2)$ spin irreps are the conjugates $[\bar{f}]_{\sigma_-}\otimes [\bar{f}]_{\sigma_+}$ from which the spins of each shell $S_{\sigma_-}\otimes S_{\sigma_+}$ can be obtained along with the total coupling to spin $S_{\sigma}=0$ by means of regular angular momentum coupling methods. The total nuclear spin $S$ is obtained straightforwardly as well.

The next step in defining the nuclear state is the reduction of the spatial algebras into trivial algebras. The proxy-$SU(3)$ requires such reduction to include $SU(3)$ which can be done by solving the branching problem $U(\Omega)\supset SU(3)$ \cite{DRAAYER1989279}. This decomposition has to be performed independently for each shell involved in the model, \textit{i.e}, $U(\Omega_{\sigma-})\otimes U(\Omega_{\sigma+})\supset \alpha_{-}SU(3)_{\sigma-}\otimes \alpha_{+}SU(3)_{\sigma+}$ where the $\alpha$ represent  multiplicities. The total nucleon irreps $\rho_\sigma(\lambda,\mu)_{\sigma}$ where $\rho_{\sigma}$ represent the multiplicities arises from the irreps in the Kronecker product $(\lambda,\mu)_{\sigma-}\otimes (\lambda,\mu)_{\sigma+}$. Then, in a similar way the total nucleus irreps $\rho(\lambda,\mu)$ are obtained from the Kronecker product $(\lambda,\mu)_\pi\otimes (\lambda,\mu)_\nu$. The physical basis decomposition $SU(3)\supset SO(3)$ is adopted in the subsequent branching in order to obtain the total orbital angular momentum $L$. Recall that the step $SU(3)\supset SO(3)$ is non-canonical so that an additional label $K$ appears that accounts for the multiplicities of $L$ in a given irrep $(\lambda,\mu)$  \cite{sloane2014lie}. This quantum number is usually called Elliott label which is used to denote rotational bands and its values are $K = \text{min}(\lambda,\mu), \text{min}(\lambda,\mu)-2, \text{min}(\lambda,\mu)-4, ..., 0$ or $1$ from which the Vergados label $\bar{\chi}$ is obtained by orthogonalization \cite{VERGADOS1968681}. The final coupling involves the product of total orbital angular momentum $L$ and total spin $S$ which form the total angular momentum $J$ and its projection $M$. 

The algebra chain and the corresponding state can be seen in equations \ref{chain} and \ref{state} respectively. For convenience reasons,  the state will be denoted sometimes as $|\Psi_{\pi+\nu}\rangle = |\Omega; J, M \rangle$ where the label $\Omega$ encompasses all quantum numbers of higher algebras than those of  total angular momentum $SU_J(2)$ and its projection $U_M(1)$.

\begin{widetext}

\begin{equation}
\resizebox{0.9\hsize}{!}{$
\begin{aligned} 
U&\left(2(\Omega_{\pi-}+\Omega_{\pi+})\right) \otimes U\left(2(\Omega_{\nu-}+\Omega_{\nu+})\right)\supset
\\&\hspace{4mm}\Bigg\{\Big\{\Big(U\left(\Omega_{\pi-}\right)\otimes U\left(\Omega_{\pi+}\right)\Big)\supset \Big(SU_{\pi-}(3)\otimes SU_{\pi+}(3)\Big)\supset  SU_{\pi}(3)\Big\}\otimes\Big\{\Big(SU_{\pi-}(2)\otimes SU_{\pi+}(2)\Big)\supset SU_{\pi}(2)\Big\}\Bigg\}
\\&\otimes\Bigg\{ \Big\{\Big(U\left(\Omega_{\nu-}\right)\otimes U\left(\Omega_{\nu+}\right)\Big)\supset \Big(SU_{\nu-}(3)\otimes SU_{\nu+}(3)\Big)\supset  SU_{\nu}(3)\Big\}\otimes\Big\{\Big(SU_{\nu-}(2)\otimes SU_{\nu+}(2)\Big)\supset SU_{\nu}(2)\Big\}\Bigg\}
\\&\supset\Bigg\{ SU(3)_{\pi+\nu}\supset SO_{\pi+\nu}(3)\Bigg\}\otimes SU_{\pi+\nu}(2)\supset SU_J(2)\supset U_M(1),
\end{aligned} 
\label{chain}
$}
\end{equation}

\begin{equation}
\resizebox{0.85\hsize}{!}{$
\begin{aligned} 
|\Psi_{\pi+\nu}\rangle& =\Bigg|\Big\{[1^{N_\pi}], [f_{\pi-}]\otimes[f_{\pi+}], \alpha_{\pi-}(\lambda,\mu)_{\pi-}\otimes \alpha_{\pi+}(\lambda,\mu)_{\pi+}; \rho_{\pi}(\lambda,\mu)_{\pi}, S_{\pi-}\otimes S_{\pi+}; S_{\pi}\Big\} \\& \hspace{3.2mm}\otimes \Big\{[1^{N_\nu}], [f_{\nu-}]\otimes[f_{\nu+}],  \alpha_{\nu-}(\lambda,\mu)_{\nu-}\otimes \alpha_{\nu+}(\lambda,\mu)_{\nu+}; \rho_{\nu}(\lambda,\mu)_{\nu}, S_{\nu-}\otimes S_{\nu+}; S_{\nu}\Big\};\\&\hspace{8mm}
\rho(\lambda,\mu), K, L, S; J, M \Bigg\rangle.
\end{aligned} 
$}
\label{state}
\end{equation}
\end{widetext}

\section{OCTUPOLE OPERATOR}

The octupole operator is defined as a sum of proton and neutron contributions 
\begin{equation}
\begin{aligned}
\mathcal{O}^{3}_{\mu} = \mathcal{O}^{\hspace{2mm}3}_{\pi\mu}+\mathcal{O}^{\hspace{2mm}3}_{\nu\mu},
\end{aligned} 
\label{operatorfull}
\end{equation}
with the single nucleon term given by
\begin{equation}
\begin{aligned}
\mathcal{O}^{\hspace{2mm}3}_{\sigma\mu}=\sum_{i=1}^{Z(N)}r^3_{\sigma}(i)Y^{3}_{\mu}(\hat{r}_\sigma(i)),
\end{aligned} 
\label{operator}
\end{equation}
where $\sigma=\pi,\nu$. Following the methodology in \cite{restrepo2024}, the mathematical form of the  operator in terms of one-body $SU(3)$ tensors is given by equation \ref{expansion} where the expansion coefficients $\mathcal{V}^{L_o=3}_{\sigma}$ are displayed in table \ref{table1}. These coefficients condense the products of $SU_J(2)$ reduced matrix elements \cite{moshinsky1996harmonic} and coupling coefficients. Notice that the sum over $\eta,\eta'$ runs over the two shells where the operator is defined; these are 5,6 and 6,7 for the proton and neutron cases respectively. 

\begin{equation}
\resizebox{0.9\hsize}{!}{$
\begin{aligned}
\mathcal{O}_{\sigma \hspace{7mm}M_o}^{\hspace{2mm}L_o=J_o=3} &= \sum_{\eta,\eta',(\lambda_o,\mu_o),K_o}\mathcal{V}^{L_o=3}_{\sigma}(\eta,\eta',\lambda_o,\mu_o,K_o)
\\&\times\big\{a^{\dagger}_{(\eta',0)\frac{1}{2}}\tilde{a}_{(0,\eta)\frac{1}{2}}\big\}_{\sigma\hspace{27mm}M_o}^{\rho_o=1,(\lambda_o,\mu_o),K_o,L_o=J_o=3}.
\end{aligned} 
$}
\label{expansion}
\end{equation}

The irrep $(\lambda_o,\mu_o)$ arises from the Kronecker product between the single particle irreps $(\eta',0)\otimes(0,\eta)$. A sum over the label $K_o$ must be explicit since the angular momentum $L_o = 3$ may have several multiplicities in a single $(\lambda_o,\mu_o)$ as indicated in table \ref{table1}. Notice that the coefficients $\mathcal{V}_{\sigma}$ of the irreps $(\lambda_o, \mu_o)$ and $(\mu_o, \lambda_o)$ are equal.

\begin{table*}[!t]
\caption{\label{table1}
Expansion coefficients
$\mathcal{V}_{\sigma}^{L_o =3}(\eta,\eta',\lambda_o,\mu_o, K_o)$ of $\mathcal{O}^{\hspace{2mm}3}_{\sigma\mu}$ for actinides.} 
\begin{ruledtabular}
\begin{tabular}{cccccccccccc}
$(\lambda_o,\mu_o)$ & \multicolumn{2}{c}{(6,7),(7,6)}& \multicolumn{2}{c}{(5,6),(6,5)}& \multicolumn{2}{c}{(4,5),(5,4)}& \multicolumn{2}{c}{(3,4),(4,3)}& \multicolumn{2}{c}{(2,3),(3,2)}& \multicolumn{1}{c}{(1,2),(2,1)}\\
\cmidrule(l){2-12}
$K_o$ & 0 & 2 & 1  & 3 & 0 & 2
& 1 & 3 & 0 & 2 & 1 \\
\hline
&&&&&&&&\\
$\mathcal{V}_{\pi}^{3}$&  &  & 9.7515 & -9.9778 & -5.1484 & 5.8693 & -7.2727 & 13.1012 & 16.0658 & -7.1607 & 0.2676 \\
$\mathcal{V}_{\nu}^{3}$& 2.2092 & -16.0225 & 1.6038 & -8.1959 & -6.6155 & 15.1021 & -2.6772 & 15.8482 & 22.1137 & -14.1431 & -5.3526  \\
\end{tabular}
\end{ruledtabular}
\end{table*}

\section{HAMILTONIAN AND MATRIX ELEMENTS}

Elliott's original model included a mean field and a residual quadrupole-quadrupole interaction in the nuclear Hamiltonian which accounted for collective excitations of nuclei. We include an octupole-octupole two-body residual term in order to consider such degree of freedom as well. This extended Hamiltonian is given by
\begin{equation}
\begin{aligned}
H = \hbar\omega\hat{n}+\chi \mathcal{Q}^2\cdot \mathcal{Q}^2 +\tau \mathcal{O}^3\cdot \mathcal{O}^3 -H_{g.s.},
\end{aligned} 
\label{hamiltonian}
\end{equation}
where the first term is the harmonic oscillator mean field with $\hat{n}$  the $U(3)$ Casimir operator of order 1 that accounts for the number of particles in the excited shell and the strength $\hbar\omega=45A^{-1/3}-25A^{-2/3}$ MeV. The second term represents the quadrupole residual interaction which in $SU(3)$ symmetry-based models is given by the relation $\mathcal{Q}^2\cdot \mathcal{Q}^2 = 4\mathcal{C}^{2}(\lambda,\mu) -3L^2$ with the $SU(3)$ Casimir operator of order two given by $\mathcal{C}^{2}(\lambda,\mu) = \lambda^2 +\mu^2 +\lambda\mu+3(\lambda + \mu)$ and the total orbital angular momentum denoted by $L$. The third term is the octupole residual interaction which is constructed as a scalar product of the expansion in equation \ref{expansion}. The last term $H_{g.s.}$ is the energy eigenvalue of the ground state which sets it to 0 MeV. The parameters $\chi$ and $\tau$ are obtained from fitting procedure to experimental energy levels.

The reduced matrix elements (RME) of the term $\mathcal{O}^3\cdot \mathcal{O}^3$ are given by equation \ref{matrixelement} where an identity $\sum_{\Psi''_{\pi+\nu}}|\Psi''_{\pi+\nu}\rangle\langle\Psi''_{\pi+\nu}|$ had to be introduced and the tensor expansion of equation \ref{expansion} was replaced explicitly. The RME on the right hand side of equation \ref{matrixelement} can be computed with equation 10 of reference \cite{TROLTENIER199553}, which separates the proton and neutron spaces in products of recoupling coefficients and triple-barred RME. The later factors however, require an additional mathematical procedure for their calculation as explained in appendix A. Computer codes were developed for the calculation of these octupole interaction double-barred RME.

\begin{widetext}
\begin{equation}
\resizebox{0.95\hsize}{!}{$
\begin{aligned}
\langle\Omega; J||\mathcal{O}^3\cdot \mathcal{O}^3||\Omega'; J'\rangle &= 
 \sum_{\substack{\sigma, \sigma'\\(\lambda_o,\mu_o),K_o,\eta,\eta'\\(\lambda_o',\mu_o'),K_o',\eta'',\eta'''\\ \Omega''; J''}}
 (-1)^{J+J''}\delta_{JJ'}\sqrt{\frac{2J+1}{2J''+1}}
 \mathcal{V}_{\sigma}(\lambda_o,\mu_o,K_o,\eta,\eta')\hspace{1mm}
 \mathcal{V}_{\sigma'}(\lambda_o',\mu_o',K_o',\eta'',\eta''')
\\ &\times \Big\langle\Omega; J\Big|\Big|
\big\{a^{\dagger}_{(\eta',0)\frac{1}{2}}\tilde{a}_{(0,\eta)\frac{1}{2}}\big\}_{\sigma}^{(\lambda_o,\mu_o),K_o,L_o=J_o=3}
\Big|\Big|\Omega''; J''\Big\rangle \Big\langle\Omega''; J''\Big|\Big| \big\{a^{\dagger}_{(\eta''',0)\frac{1}{2}}\tilde{a}_{(0,\eta'')\frac{1}{2}}\big\}_{\sigma'}^{(\lambda_o',\mu_o'),K_o',L_o=J_o=3}
\Big|\Big|\Omega'; J'\Big\rangle.
\end{aligned} 
\label{matrixelement}
$}
\end{equation}
\end{widetext}

\section{$^{224}\text{Th}$ SPECTRUM}

As stated above, the isotope $^{224}$Th is one of the best candidates for research on “pear shape" deformation. Its low energy spectrum has been studied previously in \cite{PhysRevC.101.054306} where a Hamiltonian involving the $SU(3)$ Casimir operators of orders two and three was formulated. In this section we take the Hamiltonian in \ref{hamiltonian} and use equation \ref{matrixelement} to calculate its matrix elements which are degenerate with respect to the total angular momentum projection $M$. The harmonic oscillator strength has the value of  $\hbar\omega = 6.7318$ MeV.

This isotope has 16 particles in the valence shell; 8 protons and 8 neutrons which allow for the formation of 4 quartets. Thus, its ground state is given by the spatial representation $[2^4]_{\pi_-}\otimes[0]_{\pi_+}\otimes[2^4]_{\nu_-}\otimes[0]_{\nu_+}$. The reduction $U(N)\supset SU(3)$ is given by the leading \cite{kota20203} irreps  $(26,4)_{\pi_-}\otimes(0,0)_{\pi_+}\otimes(34,4)_{\nu_-}\otimes(0,0)_{\nu_+}$ which couple to a leading $(60,8)$. The band $K^\Pi=0^+$ contained in this irrep corresponds to the ground state band observed experimentally. As explained above, the excited state has two contributions; the proton excitation given by $[2^3,1]_{\pi_-}\otimes[1]_{\pi_+}\otimes[2^4]_{\nu_-}\otimes[0]_{\nu_+}$ and the neutron excitation given by $[2^4]_{\pi_-}\otimes[0]_{\pi_+}\otimes[2^3,1]_{\nu_-}\otimes[1]_{\nu_+}$. The $SU(3)$ reductions are the leading 
$(25,2)_{\pi_-}\otimes(6,0)_{\pi_+}\otimes(34,4)_{\nu_-}\otimes(0,0)_{\nu_+}$ and $(26,4)_{\pi_-}\otimes(0,0)_{\pi_+}\otimes(32,2)_{\nu_-}\otimes(7,0)_{\nu_+}$ respectively, both of which couple to total leading $(65,6)$. The band $K^\Pi=0^-$ contained in this irrep corresponds to the octupole excited band observed experimentally.

The Hamiltonian matrix elements to be calculated are those between the states observed experimentally which can be used to fit the parameters $\chi$ and $\tau$. The ground state band has angular momentum $J^\Pi=0^+,2^+,..., 18^+$ and those of the excited band are $J^\Pi=1^-,3^-,..., 17^-$. They are a total of 28 states distinguishing between the proton and neutron excited states. The intermediate states $|\Psi''_{\pi+\nu}\rangle$ of equation \ref{matrixelement} are too many to consider them all in a realistic calculation, thus we propose a truncation of the full model space to the ground and excited bands $(60,8)0^+$ and $(65,6)0^-$. These are a total of 97 intermediate states, even though not all of them will contribute because many will cancel the one-body RME due to angular momentum selection rules. Adopting these restrictions, there will be 37 non-zero independent matrix elements of the Hamiltonian, 9 of which are non-diagonal. In this restricted space the products $\mathcal{O}^3_{\pi}\cdot\mathcal{O}^3_{\pi}$ and $\mathcal{O}^3_{\nu}\cdot\mathcal{O}^3_{\nu}$ will contribute to the diagonal terms while the $\mathcal{O}^3_{\pi}\cdot\mathcal{O}^3_{\nu}$ and $\mathcal{O}^3_{\nu}\cdot\mathcal{O}^3_{\pi}$ will form the non-diagonal matrix elements. These matrix elements are shown in appendix B.

The Elliott Hamiltonian fit to $^{224}$Th spectrum results in a quadrupole-quadrupole interaction strength of $\chi = -0.0032$ and a mean square error (MSE) of 0.1771 MeV$^2$ to the experimental energy levels. The two-parameter Hamiltonian proposed in \cite{PhysRevC.101.054306} obtains an improvement with a MSE of 0.1180 MeV$^2$. The states of the negative parity band have two contributions from the model, taken into account as a superposition 
$|\Psi_{J}\rangle = \sqrt{\alpha}|\Psi^{(\pi)}_{J}\rangle+\sqrt{1-\alpha}|\Psi^{(\nu)}_{J}\rangle$, where the proton and neutron excited states are denoted as $|\Psi^{(\pi)}_{J}\rangle$, $|\Psi^{(\nu)}_{J}\rangle$ respectively and the parameter $\alpha$ expresses the contribution of each nucleon type wavefunction to the total nuclear wavefunction. While the matrix elements of the ground state band are diagonal, the energy of the negative parity band states are be given by
\begin{equation}
\begin{aligned} 
\langle\Psi_{J}|H|\Psi_{J}\rangle &= \hbar\omega+\chi \Bigg( 4\mathcal{C}^{2}(65,6) -3L(L+1)\Bigg) \\&+\tau \Bigg(\alpha\langle\Psi_{J}^{(\pi)}|\mathcal{O}^3\cdot \mathcal{O}^3|\Psi_{J}^{(\pi)}\rangle
\\&+2\sqrt{\alpha(1-\alpha)}\langle\Psi_{J}^{(\pi)}|\mathcal{O}^3\cdot \mathcal{O}^3|\Psi_{J}^{(\nu)}\rangle
\\&+(1-\alpha)\langle\Psi_{J}^{(\nu)}|\mathcal{O}^3\cdot \mathcal{O}^3|\Psi_{J}^{(\nu)}\rangle
\Bigg) -H_{g.s}.
\end{aligned} 
\label{Eoct}
\end{equation}

\begin{figure}[t!]
\includegraphics[trim={1.5cm 0 0 0}, clip, width=0.48\textwidth]{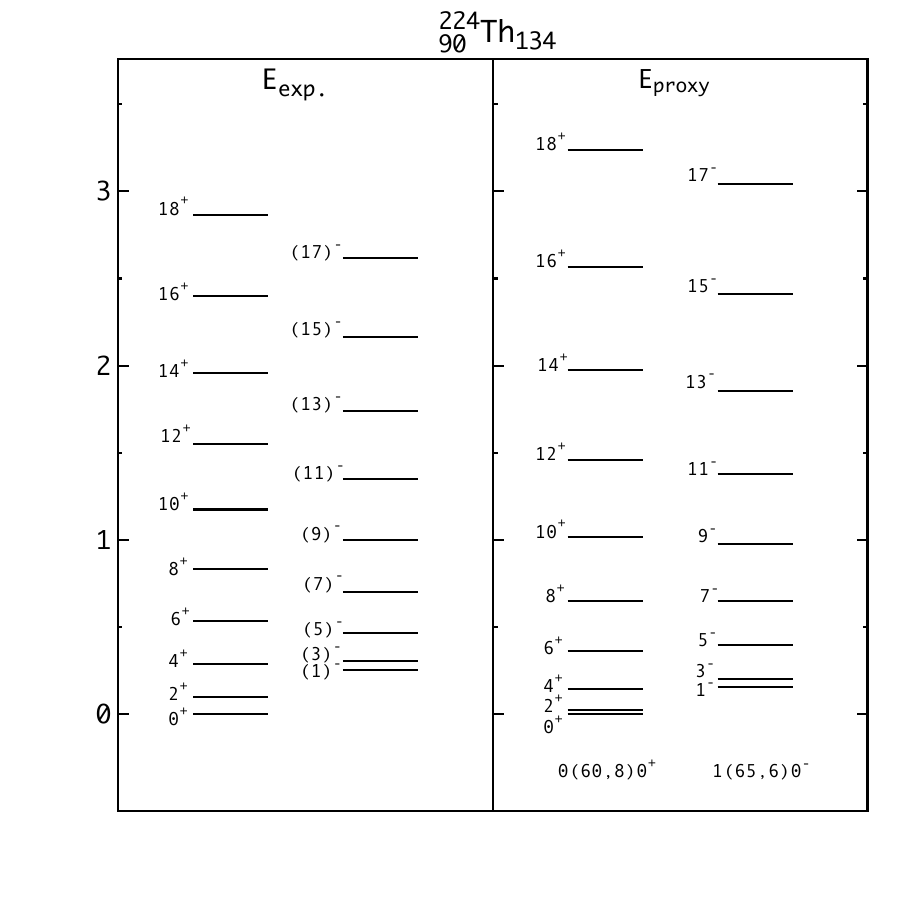}
\caption{Experimental and theoretical low energy spectra of $^{224}$Th. The energy is measured
in MeV. The labels of each band are $n(\lambda,\mu)K^{\Pi}$.} 
\label{scheme}
\end{figure}

Computer codes were developed to calculate these 
Hamiltonian expectation values and fit its three parameters resulting in $\chi = -0.0032$, $\tau = -0.0016$ and $\alpha =0.2398$. It is equivalent to a simultaneous diagonalization and fit of the parameters. This model has a MSE of 0.1739 MeV$^2$ and its spectrum is shown in figure \ref{scheme}.


\section{$B(E3; 3^-_1\rightarrow 0^+_1)$ Strength}

Using the scheme and equations shown above, we can study the $3^-_1\rightarrow 0^+_1$ electromagnetic transition strength. The electric octupole transition operator is \cite{brussaard1977shell}
\begin{equation}
\begin{aligned} 
T^3_{\mu}(E) = b_0^3\Big(e_{\pi}\mathcal{O}^3_{\mu\pi} +e_{\nu}\mathcal{O}^3_{\mu\pi}\Big),
\end{aligned} 
\label{t3operator}
\end{equation}
where the harmonic oscillator amplitude parameter is $b_0 = A^{1/6}$ fm, the proton and neutron effective charges are $e_{\pi}$ and $e_{\nu}$ respectively. The reduced transition probability is be given by  
\begin{equation}
\begin{aligned} 
B(E3; 3^-_1\rightarrow 0^+_1) = \frac{1}{7}|\langle\Psi_{J=0}||\hspace{1mm} T^3(E) \hspace{1mm}||\Psi'_{J=3}\rangle|^2,
\end{aligned} 
\label{BE3}
\end{equation}
with the reduced matrix element 
\begin{equation}
\begin{aligned} 
\langle\Psi_{J=0}||\hspace{1mm} T^3(E) \hspace{1mm}||\Psi'_{J=3}\rangle = &b_0^3\Big(a\langle\Psi_{J=0}||\hspace{1mm}\mathcal{O}^3_{\pi} \hspace{1mm}||\Psi'^{(\pi)}_{J=3}\rangle \\+&b\langle\Psi_{J=0}||\hspace{1mm}\mathcal{O}^3_{\nu}\hspace{1mm}||\Psi'^{(\nu)}_{J=3}\rangle\Big),
\end{aligned} 
\label{RMETE3}
\end{equation}
where $|\Psi^{(\sigma)}_{J=3}\rangle$ represents the excited state of nucleon $\sigma$ according to the scheme of equation \ref{state}, parameters $a$, $b$ involve the effective charges and superposition parameter $\alpha$ in the total excited state $|\Psi_{J=3}\rangle$ as $a = e_{\pi}\sqrt{\alpha}$ and $b = e_{\nu}\sqrt{1-\alpha}$ \cite{restrepo2024}. Recall that the latest evaluated experimental values of the transition $3^-_1\rightarrow 0^+_1$ are shown in figure \ref{BE3N}

From the collective model the following relation can be demonstrated \cite{Butler_2016, LEANDER1984375}
\begin{equation}
B(E3; 3^-_1\rightarrow 0^+_1) = \frac{1}{7}\Bigg(\frac{3ZeR^3\beta_3}{4\pi}\Bigg)^2,
\label{BE3Collective}
\end{equation}
using the experimental value for quadrupole deformation $\beta_2=0.174(6)$ \cite{nndc}, the calculation for octupole deformation $\beta_3=-0.153$ \cite{MOLLER20161} for $^{224}$Th and the approximation for the nuclear radius $R\approx 1.2A^{1/3}$ fm, results in a transition strength of 77.6001 W.u. This value is close to the one reported in \cite{PhysRevC.88.051302, gaffney} by different methods. 

Assuming $b\approx-a/2$ according to the results in \cite{restrepo2024} and considering the collective transition strength, one can estimate the value of $a\approx0.7805e$. Under these assumptions and the value for $\alpha$ obtained in the previous section, minimums for $e_\pi$ and $e_\nu$ are set in the model. A direct measurement of the $B(E3; 3^-_1\rightarrow 0^+_1)$ transition for $^{224}$Th is required for further analysis.

\section{CONCLUSIONS}
We obtained the mathematical form of the octupole operator $r^3 Y^3_\mu(\Omega)$ in terms of $SU(3)$ tensors for the actinide region which is minimally defined in the shells 5,6 in the proton space and 6,7 in the neutron space. The scalar operator $\mathcal{O}^3\cdot\mathcal{O}^3$ is constructed from this expansion and its reduced matrix elements in the two-shell states formalism are computed in terms of products of recoupling coefficients and single-shell reduced matrix elements. Computer codes were developed that integrate several $SU(3)$ libraries \cite{schurprogram, DYTRYCH2021108137, BAHRI199459, DRAAYER1989279} in order to perform the calculations of this work. These are available under request to the authors. Essential algebraic expressions for this work are presented in appendix A along with identities used to check the computer codes developed. 

The Hamiltonian proposed carries three parameters $\chi, \tau$ and $\alpha$ that account for the strength of the quadrupole, octupole residual interactions and nucleon contributions respectively. This model was applied to the low-lying spectrum of the isotope $^{224}$Th obtaining a slight improvement with respect to the original Elliott model. Increasing the number of parameters of the model would improve the fit but decrease its explanatory capabilities. A possible improvement might be achieved by a weaker truncation of the intermediate states in equation \ref{matrixelement}, for the influence of the octupole-octupole interaction is too little in this reduced space. An analysis on the transition $B(E3; 3^-_1\rightarrow 0^+_1)$ involving previous works on collective models, density functional theory and the scheme in this article allows to estimate the octupole effective charges of the model under certain assumptions, however, the lack of experimental data is challenging for this theoretical aspect. We hope to motivate research on this isotope from the experimental side as well.

 The octupole deformation of nuclei is a research area that will help physicists obtain a better understanding of the nuclear structure and gain insight of the fundamental symmetries of nature. With this work, we expect to increase interest in this topic from the $SU(3)$-symmetry based models and await the results of experimental efforts from novel radioactive beams accelerators which will allow the exploration of previously inaccessible nuclei.

\begin{acknowledgments}
We would like to thank professors J. G. Hirsch, V. K. B. Kota, D. Langr and J. Cseh for very valuable discussions. We thank the Institute of Physics at University of Antioquia for supporting this research.
\end{acknowledgments}

\appendix

\section{
$SU(3)$ one-body triple-barred reduced matrix elements in two-shell states}

The one-body double-barred reduced matrix elements (RME) of equation \ref{matrixelement} can be computed with equation 10 of reference \cite{TROLTENIER199553}, which separates the proton an neutron spaces obtaining a mathematical expression in terms of recoupling coefficients and triple-barred RME. The former can be computed by $SU(3)$ libraries \cite{DYTRYCH2021108137} while the latter require an additional mathematical calculation.

The proton and neutron spaces are formed each by two shells; a lower shell denoted by sub-index $-$ and an upper one denoted by  sub-index $+$, with the creation an destruction operators acting on different quantum numbers. A separation of these two shells \cite{Millener1991SU3IS} can be performed as shown in equation \ref{recouplingOB} where two situations can happen; $a^\dagger$ creates a particle in the $-$ shell and $\tilde{a}$ destroys a particle in $+$ shell, or $a^\dagger$ creates a particle in the $+$ shell and $\tilde{a}$ destroys a particle in shell $-$. The first case is a de-excitation of the nucleus while the second is an excitation process. These two cases will imply a minimal change in their mathematical expressions which are distinguished with the labels $s$ and $t$ as follows: $s = +$ and  $t=-$ identify the de-excitation process and $s = -$ and  $t=+$ identifies the excitation process as indicated in \ref{recouplingOB}. The triple-barred RME appearing on the right hand side of equation \ref{recouplingOB} are in a single shell which can be evaluated using computer packages \cite{BAHRI199459,J2007ReducedME}. A computer program was developed to perform the calculation of equation \ref{recouplingOB} depending on the case as explained above. The following relation is useful 
\begin{equation}
\resizebox{0.95\hsize}{!}{$
\begin{aligned}
&\Bigg(\big\{a^{\dagger}_{(\eta'0)\frac{1}{2}}\tilde{a}_{(0\eta)\frac{1}{2}}\big\}^{\rho_0=1,(\lambda_0\mu_0),K_o,L_o,S_0}_{\hspace{16mm}M_{L_o},M_{S_o}}\Bigg)^\dagger
\\ &= (-1)^{\eta-\eta'-L_o-M_{L_o}-M_{S_o}} \big\{a^{\dagger}_{(\eta0)\frac{1}{2}}\tilde{a}_{(0\eta')\frac{1}{2}}\big\}^{\rho_0=1,(\lambda_0\mu_0),K_o,L_o,S_0}_{\hspace{13mm}-M_{L_o},-M_{S_o}},\\
\end{aligned} 
\label{dagger}
$}
\end{equation}
and the next identities \cite{suhonen2007nucleons, VERGADOS1968681} must be hold which were used to check the computer codes

\begin{equation}
\resizebox{0.9\hsize}{!}{$
\begin{aligned}
&\Big\langle L', S'\Big|\Big| T^{L_o, S_o} \Big|\Big| L,S\Big\rangle 
\\&= (-1)^{L-L'+S-S'}\sqrt{\frac{(2L+1)(2S+1)}{(2L'+1)(2S'+1)}}\Big\langle L,S\Big|\Big|T^{L_o, S_o}  \Big|\Big| L',S'\Big\rangle^*,
\end{aligned} 
\label{identity3b}
$}
\end{equation}

\begin{equation}
\resizebox{0.78\hsize}{!}{$
\begin{aligned}
&\frac{\langle(\lambda_1,\mu_1)|||T^{(\mu_2,\lambda_2)} |||(\lambda_3,\mu_3)\rangle}{\langle(\lambda_3,\mu_3)|||T^{(\lambda_2,\mu_2)} |||(\lambda_1,\mu_1)\rangle}  \\ &=(-1)^{\frac{1}{2}(\mu_1+\mu_2-\mu_3-\lambda_1-\lambda_2+\lambda_3)}\sqrt{\frac{\text{dim}(\lambda_3,\mu_3)}{\text{dim}(\lambda_1,\mu_1)}}.
\end{aligned} 
\label{identity3b}
$}
\end{equation}

\begin{widetext}
\begin{equation}
\resizebox{0.93\hsize}{!}{$
\begin{aligned} 
&\Big\langle[f'_-]\otimes[f'_+],(\lambda'_-\mu'_-)\otimes(\lambda'_+\mu'_+),\rho'(\lambda'\mu'),S'\Big|\Big|\Big| \big\{a^{\dagger}_{(\eta'0)\frac{1}{2}}\tilde{a}_{(0\eta)\frac{1}{2}}\big\}^{\rho_0=1,(\lambda_0\mu_0)S_0}\Big|\Big|\Big|[f_-]\otimes[f_+],(\lambda_-\mu_-)\otimes(\lambda_+\mu_+),\rho(\lambda\mu),S\Big\rangle_{\bar{\rho}}
\\& \hspace{20mm} = \begin{Bmatrix}
  (\lambda_-,\mu_-) & (\alpha, \beta) & (\lambda'_-,\mu'_-) & \rho=1\\
  (\lambda_+,\mu_+) & (\gamma,\delta) & (\lambda'_+,\mu'_+) & \rho=1\\
  (\lambda,\mu) & (\lambda_o,\mu_o) & (\lambda',\mu') & \bar{\rho}\\
  \rho & \rho=1 & \rho'&
 \end{Bmatrix}
\chi\begin{Bmatrix}
  S_- & \frac{1}{2} & S'_- \\
  S_+ & \frac{1}{2} & S'_+ \\
  S & S_o & S' \\
 \end{Bmatrix}(-1)^{\phi}\sqrt{\frac{\text{dim}(\lambda_s\mu_s)}{\text{dim}(\lambda'_s\mu'_s)}}\frac{\sqrt{2S_s+1}}{\sqrt{2S'_s+1}}
\\\\&\hspace{20mm} \times\Big\langle[f'_t],(\lambda'_t\mu'_t),S'_t\Big|\Big|\Big|a^{\dagger}_{(\eta'0)\frac{1}{2}}\Big|\Big|\Big| [f_t],(\lambda_t\mu_t)S_t\Big\rangle_{\rho=1}
\Big\langle[f_s],(\lambda_s\mu_s),S_s\Big|\Big|\Big|a^{\dagger}_{(\eta0)\frac{1}{2}}\Big|\Big|\Big|[f'_s],(\lambda'_s\mu'_s),S'_s\Big\rangle_{\rho=1},
\end{aligned} 
\label{recouplingOB}
$}
\end{equation}
$$
\begin{cases}
    \phi = \eta + \frac{1}{2} +S_+-S'_++\lambda'_+-\lambda_++\mu'_+-\mu_+,\\\hspace{10mm}(\alpha,\beta)=(\eta',0)\hspace{1mm}\text{and}\hspace{1mm}  (\gamma,\delta)=(0,\eta) & \text{if $s=+$ and $t=-$}\\ \\
    \phi = \eta' + \frac{3}{2} +S_--S'_--S_o+\lambda'_--\lambda_-+\mu'_--\mu_--\lambda_o-\mu_o,\\\hspace{10mm}(\alpha,\beta)=(0,\eta)\hspace{1mm}\text{and}\hspace{1mm}  (\gamma,\delta)=(\eta',0) & \text{if $s=-$ and $t=+$ }.\\		
\end{cases}
$$
\end{widetext}

\section{$\mathcal{O}\cdot\mathcal{O}$ matrix elements}

In this section we present the matrix elements of the operator $\mathcal{O}\cdot\mathcal{O}$ in table \ref{OOmatrix} as defined in equation $\ref{matrixelement}$.

\onecolumngrid
\vspace*{\fill}
\begin{sidewaystable}[p!]
\begin{center}
\centering
\caption{\label{OOmatrix}
Reduced matrix elements of $\mathcal{O}\cdot\mathcal{O}$. The notation of the states $J^\Pi_\sigma$ indicates the angular momentum, parity and the excited nucleon $\sigma$. In the ground state band, this last label is omitted.}
 \resizebox{\textwidth}{!}{%
\begin{tabular}{c|cccccccccccccccccccccccccccc}
 & $0^+$ & $2^+$ & $4^+$ & $6^+$ & $8^+$ & $10^+$ & $12^+$ & $14^+$ & $16^+$ & $18^+$ & $1_{\pi}^-$ & $3_{\pi}^-$ & $5_{\pi}^-$ & $7_{\pi}^-$  & $9_{\pi}^-$ & $11_{\pi}^-$ & $13_{\pi}^-$ & $15_{\pi}^-$ & $17_{\pi}^-$ & $1_{\nu}^-$ & $3_{\nu}^-$ & $5_{\nu}^-$ & $7_{\nu}^-$  & $9_{\nu}^-$ & $11_{\nu}^-$ & $13_{\nu}^-$ & $15_{\nu}^-$ & $17_{\nu}^-$ \\
\hline
$0^+$ & 4.5946 & 0 & 0 & 0 & 0 & 0 & 0 & 0 & 0 & 0 & 0 & 0 & 0 & 0 & 0 & 0 & 0 & 0 & 0 & 0 & 0 & 0 & 0 & 0 & 0 & 0 & 0 & 0 \\
$2^+$ & 0 & 25.8974 & 0 & 0 & 0 & 0 & 0 & 0 & 0 & 0 & 0 & 0 & 0 & 0 & 0 & 0 & 0 & 0 & 0 & 0 & 0 & 0 & 0 & 0 & 0 & 0 & 0 & 0 \\
$4^+$ & 0 & 0 & 36.6437 & 0 & 0 & 0 & 0 & 0 & 0 & 0 & 0 & 0 & 0 & 0 & 0 & 0 & 0 & 0 & 0 & 0 & 0 & 0 & 0 & 0 & 0 & 0 & 0 & 0 \\
$6^+$ & 0 & 0 & 0 & 32.0281 & 0 & 0 & 0 & 0 & 0 & 0 & 0 & 0 & 0 & 0 & 0 & 0 & 0 & 0 & 0 & 0 & 0 & 0 & 0 & 0 & 0 & 0 & 0 & 0 \\
$8^+$ & 0 & 0 & 0 & 0 & 30.4360 & 0 & 0 & 0 & 0 & 0 & 0 & 0 & 0 & 0 & 0 & 0 & 0 & 0 & 0 & 0 & 0 & 0 & 0 & 0 & 0 & 0 & 0 & 0 \\
$10^+$ & 0 & 0 & 0 & 0 & 0 & 29.4780 & 0 & 0 & 0 & 0 & 0 & 0 & 0 & 0 & 0 & 0 & 0 & 0 & 0 & 0 & 0 & 0 & 0 & 0 & 0 & 0 & 0 & 0 \\
$12^+$ & 0 & 0 & 0 & 0 & 0 & 0 & 28.8096 & 0 & 0 & 0 & 0 & 0 & 0 & 0 & 0 & 0 & 0 & 0 & 0 & 0 & 0 & 0 & 0 & 0 & 0 & 0 & 0 & 0 \\
$14^+$ & 0 & 0 & 0 & 0 & 0 & 0 & 0 & 28.3865 & 0 & 0 & 0 & 0 & 0 & 0 & 0 & 0 & 0 & 0 & 0 & 0 & 0 & 0 & 0 & 0 & 0 & 0 & 0 & 0 \\
$16^+$ & 0 & 0 & 0 & 0 & 0 & 0 & 0 & 0 & 28.2491 & 0 & 0 & 0 & 0 & 0 & 0 & 0 & 0 & 0 & 0 & 0 & 0 & 0 & 0 & 0 & 0 & 0 & 0 & 0 \\
$18^+$ & 0 & 0 & 0 & 0 & 0 & 0 & 0 & 0 & 0 & 28.4693 & 0 & 0 & 0 & 0 & 0 & 0 & 0 & 0 & 0 & 0 & 0 & 0 & 0 & 0 & 0 & 0 & 0 & 0 \\
$1_{\pi}^-$ & 0 & 0 & 0 & 0 & 0 & 0 & 0 & 0 & 0 & 0 & 0.0583 & 0 & 0 & 0 & 0 & 0 & 0 & 0 & 0 & 0.8057 & 0 & 0 & 0 & 0 & 0 & 0 & 0 & 0 \\
$3_{\pi}^-$ & 0 & 0 & 0 & 0 & 0 & 0 & 0 & 0 & 0 & 0 & 0 & 0.8027 & 0 & 0 & 0 & 0 & 0 & 0 & 0 & 0 & 5.8653 & 0 & 0 & 0 & 0 & 0 & 0 & 0 \\
$5_{\pi}^-$ & 0 & 0 & 0 & 0 & 0 & 0 & 0 & 0 & 0 & 0 & 0 & 0 & 0.7432 & 0 & 0 & 0 & 0 & 0 & 0 & 0 & 0 & 3.9895 & 0 & 0 & 0 & 0 & 0 & 0 \\
$7_{\pi}^-$ & 0 & 0 & 0 & 0 & 0 & 0 & 0 & 0 & 0 & 0 & 0 & 0 & 0 & 1.0362 & 0 & 0 & 0 & 0 & 0 & 0 & 0 & 0 & 4.1776 & 0 & 0 & 0 & 0 & 0 \\
$9_{\pi}^-$ & 0 & 0 & 0 & 0 & 0 & 0 & 0 & 0 & 0 & 0 & 0 & 0 & 0 & 0 & 1.4556 & 0 & 0 & 0 & 0 & 0 & 0 & 0 & 0 & 4.6520 & 0 & 0 & 0 & 0 \\
$11_{\pi}^-$ & 0 & 0 & 0 & 0 & 0 & 0 & 0 & 0 & 0 & 0 & 0 & 0 & 0 & 0 & 0 & 1.9908 & 0 & 0 & 0 & 0 & 0 & 0 & 0 & 0 & 5.2846 & 0 & 0 & 0 \\
$13_{\pi}^-$ & 0 & 0 & 0 & 0 & 0 & 0 & 0 & 0 & 0 & 0 & 0 & 0 & 0 & 0 & 0 & 0 & 2.6439 & 0 & 0 & 0 & 0 & 0 & 0 & 0 & 0 & 6.0400 & 0 & 0 \\
$15_{\pi}^-$ & 0 & 0 & 0 & 0 & 0 & 0 & 0 & 0 & 0 & 0 & 0 & 0 & 0 & 0 & 0 & 0 & 0 & 3.4203 & 0 & 0 & 0 & 0 & 0 & 0 & 0 & 0 & 6.9057 & 0 \\
$17_{\pi}^-$ & 0 & 0 & 0 & 0 & 0 & 0 & 0 & 0 & 0 & 0 & 0 & 0 & 0 & 0 & 0 & 0 & 0 & 0 & 4.3272 & 0 & 0 & 0 & 0 & 0 & 0 & 0 & 0 & 7.8818 \\
$1_{\nu}^-$ & 0 & 0 & 0 & 0 & 0 & 0 & 0 & 0 & 0 & 0 & 0.8057 & 0 & 0 & 0 & 0 & 0 & 0 & 0 & 0 & 12.7929 & 0 & 0 & 0 & 0 & 0 & 0 & 0 & 0 \\
$3_{\nu}^-$ & 0 & 0 & 0 & 0 & 0 & 0 & 0 & 0 & 0 & 0 & 0 & 5.8653 & 0 & 0 & 0 & 0 & 0 & 0 & 0 & 0 & 50.6621 & 0 & 0 & 0 & 0 & 0 & 0 & 0 \\
$5_{\nu}^-$ & 0 & 0 & 0 & 0 & 0 & 0 & 0 & 0 & 0 & 0 & 0 & 0 & 3.9895 & 0 & 0 & 0 & 0 & 0 & 0 & 0 & 0 & 32.9312 & 0 & 0 & 0 & 0 & 0 & 0 \\
$7_{\nu}^-$ & 0 & 0 & 0 & 0 & 0 & 0 & 0 & 0 & 0 & 0 & 0 & 0 & 0 & 4.1776 & 0 & 0 & 0 & 0 & 0 & 0 & 0 & 0 & 29.9974 & 0 & 0 & 0 & 0 & 0 \\
$9_{\nu}^-$ & 0 & 0 & 0 & 0 & 0 & 0 & 0 & 0 & 0 & 0 & 0 & 0 & 0 & 0 & 4.6520 & 0 & 0 & 0 & 0 & 0 & 0 & 0 & 0 & 28.2286 & 0 & 0 & 0 & 0 \\
$11_{\nu}^-$ & 0 & 0 & 0 & 0 & 0 & 0 & 0 & 0 & 0 & 0 & 0 & 0 & 0 & 0 & 0 & 5.2846 & 0 & 0 & 0 & 0 & 0 & 0 & 0 & 0 & 26.7378 & 0 & 0 & 0 \\
$13_{\nu}^-$ & 0 & 0 & 0 & 0 & 0 & 0 & 0 & 0 & 0 & 0 & 0 & 0 & 0 & 0 & 0 & 0 & 6.0400 & 0 & 0 & 0 & 0 & 0 & 0 & 0 & 0 & 25.3676 & 0 & 0 \\
$15_{\nu}^-$ & 0 & 0 & 0 & 0 & 0 & 0 & 0 & 0 & 0 & 0 & 0 & 0 & 0 & 0 & 0 & 0 & 0 & 6.9057 & 0 & 0 & 0 & 0 & 0 & 0 & 0 & 0 & 24.1192 & 0 \\
$17_{\nu}^-$ & 0 & 0 & 0 & 0 & 0 & 0 & 0 & 0 & 0 & 0 & 0 & 0 & 0 & 0 & 0 & 0 & 0 & 0 & 7.8818 & 0 & 0 & 0 & 0 & 0 & 0 & 0 & 0 & 23.0470 \\
\end{tabular}
}
\end{center}
\end{sidewaystable}
\vspace*{\fill}
\twocolumngrid












\nocite{*}

\bibliography{bib}

\providecommand{\noopsort}[1]{}\providecommand{\singleletter}[1]{#1}%
\begin{thebibliography}{44}%
\makeatletter
\providecommand \@ifxundefined [1]{%
 \@ifx{#1\undefined}
}%
\providecommand \@ifnum [1]{%
 \ifnum #1\expandafter \@firstoftwo
 \else \expandafter \@secondoftwo
 \fi
}%
\providecommand \@ifx [1]{%
 \ifx #1\expandafter \@firstoftwo
 \else \expandafter \@secondoftwo
 \fi
}%
\providecommand \natexlab [1]{#1}%
\providecommand \enquote  [1]{``#1''}%
\providecommand \bibnamefont  [1]{#1}%
\providecommand \bibfnamefont [1]{#1}%
\providecommand \citenamefont [1]{#1}%
\providecommand \href@noop [0]{\@secondoftwo}%
\providecommand \href [0]{\begingroup \@sanitize@url \@href}%
\providecommand \@href[1]{\@@startlink{#1}\@@href}%
\providecommand \@@href[1]{\endgroup#1\@@endlink}%
\providecommand \@sanitize@url [0]{\catcode `\\12\catcode `\$12\catcode `\&12\catcode `\#12\catcode `\^12\catcode `\_12\catcode `\%12\relax}%
\providecommand \@@startlink[1]{}%
\providecommand \@@endlink[0]{}%
\providecommand \url  [0]{\begingroup\@sanitize@url \@url }%
\providecommand \@url [1]{\endgroup\@href {#1}{\urlprefix }}%
\providecommand \urlprefix  [0]{URL }%
\providecommand \Eprint [0]{\href }%
\providecommand \doibase [0]{https://doi.org/}%
\providecommand \selectlanguage [0]{\@gobble}%
\providecommand \bibinfo  [0]{\@secondoftwo}%
\providecommand \bibfield  [0]{\@secondoftwo}%
\providecommand \translation [1]{[#1]}%
\providecommand \BibitemOpen [0]{}%
\providecommand \bibitemStop [0]{}%
\providecommand \bibitemNoStop [0]{.\EOS\space}%
\providecommand \EOS [0]{\spacefactor3000\relax}%
\providecommand \BibitemShut  [1]{\csname bibitem#1\endcsname}%
\let\auto@bib@innerbib\@empty
\bibitem [{\citenamefont {Heyde}\ and\ \citenamefont {Wood}(2016)}]{Heyde_2016}%
  \BibitemOpen
  \bibfield  {author} {\bibinfo {author} {\bibfnamefont {K.}~\bibnamefont {Heyde}}\ and\ \bibinfo {author} {\bibfnamefont {J.~L.}\ \bibnamefont {Wood}},\ }\bibfield  {title} {\bibinfo {title} {Nuclear shapes: from earliest ideas to multiple shape coexisting structures},\ }\href {https://doi.org/10.1088/0031-8949/91/8/083008} {\bibfield  {journal} {\bibinfo  {journal} {Physica Scripta}\ }\textbf {\bibinfo {volume} {91}},\ \bibinfo {pages} {083008} (\bibinfo {year} {2016})}\BibitemShut {NoStop}%
\bibitem [{\citenamefont {Mackintosh}(1977)}]{RSMackintosh_1977}%
  \BibitemOpen
  \bibfield  {author} {\bibinfo {author} {\bibfnamefont {R.~S.}\ \bibnamefont {Mackintosh}},\ }\bibfield  {title} {\bibinfo {title} {The shape of nuclei},\ }\href {https://doi.org/10.1088/0034-4885/40/7/001} {\bibfield  {journal} {\bibinfo  {journal} {Reports on Progress in Physics}\ }\textbf {\bibinfo {volume} {40}},\ \bibinfo {pages} {731} (\bibinfo {year} {1977})}\BibitemShut {NoStop}%
\bibitem [{\citenamefont {Nilsson}\ and\ \citenamefont {Ragnarsson}(1995)}]{nilsson1995shapes}%
  \BibitemOpen
  \bibfield  {author} {\bibinfo {author} {\bibfnamefont {S.}~\bibnamefont {Nilsson}}\ and\ \bibinfo {author} {\bibfnamefont {I.}~\bibnamefont {Ragnarsson}},\ }\href@noop {} {\emph {\bibinfo {title} {Shapes and Shells in Nuclear Structure}}}\ (\bibinfo  {publisher} {Cambridge University Press},\ \bibinfo {year} {1995})\BibitemShut {NoStop}%
\bibitem [{\citenamefont {Tajima}\ and\ \citenamefont {Suzuki}(2001)}]{PhysRevC.64.037301}%
  \BibitemOpen
  \bibfield  {author} {\bibinfo {author} {\bibfnamefont {N.}~\bibnamefont {Tajima}}\ and\ \bibinfo {author} {\bibfnamefont {N.}~\bibnamefont {Suzuki}},\ }\bibfield  {title} {\bibinfo {title} {Prolate dominance of nuclear shape caused by a strong interference between the effects of spin-orbit and ${l}^{2}$ terms of the nilsson potential},\ }\href {https://doi.org/10.1103/PhysRevC.64.037301} {\bibfield  {journal} {\bibinfo  {journal} {Phys. Rev. C}\ }\textbf {\bibinfo {volume} {64}},\ \bibinfo {pages} {037301} (\bibinfo {year} {2001})}\BibitemShut {NoStop}%
\bibitem [{\citenamefont {Heyde}\ and\ \citenamefont {Wood}(2011)}]{RevModPhys.83.1467}%
  \BibitemOpen
  \bibfield  {author} {\bibinfo {author} {\bibfnamefont {K.}~\bibnamefont {Heyde}}\ and\ \bibinfo {author} {\bibfnamefont {J.~L.}\ \bibnamefont {Wood}},\ }\bibfield  {title} {\bibinfo {title} {Shape coexistence in atomic nuclei},\ }\href {https://doi.org/10.1103/RevModPhys.83.1467} {\bibfield  {journal} {\bibinfo  {journal} {Rev. Mod. Phys.}\ }\textbf {\bibinfo {volume} {83}},\ \bibinfo {pages} {1467} (\bibinfo {year} {2011})}\BibitemShut {NoStop}%
\bibitem [{\citenamefont {Martinou}\ \emph {et~al.}(2021)\citenamefont {Martinou}, \citenamefont {Bonatsos}, \citenamefont {Mertzimekis}, \citenamefont {Karakatsanis}, \citenamefont {Assimakis}, \citenamefont {Peroulis}, \citenamefont {Sarantopoulou},\ and\ \citenamefont {Minkov}}]{Martinou2021}%
  \BibitemOpen
  \bibfield  {author} {\bibinfo {author} {\bibfnamefont {A.}~\bibnamefont {Martinou}}, \bibinfo {author} {\bibfnamefont {D.}~\bibnamefont {Bonatsos}}, \bibinfo {author} {\bibfnamefont {T.~J.}\ \bibnamefont {Mertzimekis}}, \bibinfo {author} {\bibfnamefont {K.~E.}\ \bibnamefont {Karakatsanis}}, \bibinfo {author} {\bibfnamefont {I.~E.}\ \bibnamefont {Assimakis}}, \bibinfo {author} {\bibfnamefont {S.~K.}\ \bibnamefont {Peroulis}}, \bibinfo {author} {\bibfnamefont {S.}~\bibnamefont {Sarantopoulou}},\ and\ \bibinfo {author} {\bibfnamefont {N.}~\bibnamefont {Minkov}},\ }\bibfield  {title} {\bibinfo {title} {The islands of shape coexistence within the elliott and the {proxy-SU(3}) models},\ }\href@noop {} {\bibfield  {journal} {\bibinfo  {journal} {The European Physical Journal A}\ }\textbf {\bibinfo {volume} {57}},\ \bibinfo {pages} {84} (\bibinfo {year} {2021})}\BibitemShut {NoStop}%
\bibitem [{\citenamefont {Stone}(2016)}]{STONE20161}%
  \BibitemOpen
  \bibfield  {author} {\bibinfo {author} {\bibfnamefont {N.}~\bibnamefont {Stone}},\ }\bibfield  {title} {\bibinfo {title} {Table of nuclear electric quadrupole moments},\ }\href {https://doi.org/https://doi.org/10.1016/j.adt.2015.12.002} {\bibfield  {journal} {\bibinfo  {journal} {Atomic Data and Nuclear Data Tables}\ }\textbf {\bibinfo {volume} {111-112}},\ \bibinfo {pages} {1} (\bibinfo {year} {2016})}\BibitemShut {NoStop}%
\bibitem [{\citenamefont {Bonatsos}\ \emph {et~al.}(2017{\natexlab{a}})\citenamefont {Bonatsos}, \citenamefont {Assimakis}, \citenamefont {Minkov}, \citenamefont {Martinou}, \citenamefont {Sarantopoulou}, \citenamefont {Cakirli}, \citenamefont {Casten},\ and\ \citenamefont {Blaum}}]{PhysRevC.95.064326}%
  \BibitemOpen
  \bibfield  {author} {\bibinfo {author} {\bibfnamefont {D.}~\bibnamefont {Bonatsos}}, \bibinfo {author} {\bibfnamefont {I.~E.}\ \bibnamefont {Assimakis}}, \bibinfo {author} {\bibfnamefont {N.}~\bibnamefont {Minkov}}, \bibinfo {author} {\bibfnamefont {A.}~\bibnamefont {Martinou}}, \bibinfo {author} {\bibfnamefont {S.}~\bibnamefont {Sarantopoulou}}, \bibinfo {author} {\bibfnamefont {R.~B.}\ \bibnamefont {Cakirli}}, \bibinfo {author} {\bibfnamefont {R.~F.}\ \bibnamefont {Casten}},\ and\ \bibinfo {author} {\bibfnamefont {K.}~\bibnamefont {Blaum}},\ }\bibfield  {title} {\bibinfo {title} {Analytic predictions for nuclear shapes, prolate dominance, and the prolate-oblate shape transition in the proxy-su(3) model},\ }\href {https://doi.org/10.1103/PhysRevC.95.064326} {\bibfield  {journal} {\bibinfo  {journal} {Phys. Rev. C}\ }\textbf {\bibinfo {volume} {95}},\ \bibinfo {pages} {064326} (\bibinfo {year} {2017}{\natexlab{a}})}\BibitemShut {NoStop}%
\bibitem [{\citenamefont {Butler}(2016)}]{Butler_2016}%
  \BibitemOpen
  \bibfield  {author} {\bibinfo {author} {\bibfnamefont {P.~A.}\ \bibnamefont {Butler}},\ }\bibfield  {title} {\bibinfo {title} {Octupole collectivity in nuclei},\ }\href {https://doi.org/10.1088/0954-3899/43/7/073002} {\bibfield  {journal} {\bibinfo  {journal} {Journal of Physics G: Nuclear and Particle Physics}\ }\textbf {\bibinfo {volume} {43}},\ \bibinfo {pages} {073002} (\bibinfo {year} {2016})}\BibitemShut {NoStop}%
\bibitem [{\citenamefont {Ahmad}\ and\ \citenamefont {Butler}(1993)}]{annurev}%
  \BibitemOpen
  \bibfield  {author} {\bibinfo {author} {\bibfnamefont {I.}~\bibnamefont {Ahmad}}\ and\ \bibinfo {author} {\bibfnamefont {P.~A.}\ \bibnamefont {Butler}},\ }\bibfield  {title} {\bibinfo {title} {Octupole shapes in nuclei},\ }\href {https://doi.org/https://doi.org/10.1146/annurev.ns.43.120193.000443} {\bibfield  {journal} {\bibinfo  {journal} {Annual Review of Nuclear and Particle Science}\ }\textbf {\bibinfo {volume} {43}},\ \bibinfo {pages} {71} (\bibinfo {year} {1993})}\BibitemShut {NoStop}%
\bibitem [{\citenamefont {Butler}\ and\ \citenamefont {Nazarewicz}(1996)}]{RevModPhys.68.349}%
  \BibitemOpen
  \bibfield  {author} {\bibinfo {author} {\bibfnamefont {P.~A.}\ \bibnamefont {Butler}}\ and\ \bibinfo {author} {\bibfnamefont {W.}~\bibnamefont {Nazarewicz}},\ }\bibfield  {title} {\bibinfo {title} {Intrinsic reflection asymmetry in atomic nuclei},\ }\href {https://doi.org/10.1103/RevModPhys.68.349} {\bibfield  {journal} {\bibinfo  {journal} {Rev. Mod. Phys.}\ }\textbf {\bibinfo {volume} {68}},\ \bibinfo {pages} {349} (\bibinfo {year} {1996})}\BibitemShut {NoStop}%
\bibitem [{\citenamefont {Butler}(2020)}]{butlerpear}%
  \BibitemOpen
  \bibfield  {author} {\bibinfo {author} {\bibfnamefont {P.~A.}\ \bibnamefont {Butler}},\ }\bibfield  {title} {\bibinfo {title} {Pear-shaped atomic nuclei},\ }\href {https://doi.org/10.1098/rspa.2020.0202} {\bibfield  {journal} {\bibinfo  {journal} {Proceedings of the Royal Society A: Mathematical, Physical and Engineering Sciences}\ }\textbf {\bibinfo {volume} {476}},\ \bibinfo {pages} {20200202} (\bibinfo {year} {2020})}\BibitemShut {NoStop}%
\bibitem [{\citenamefont {Pancholi}(2020)}]{pancholi2020pear}%
  \BibitemOpen
  \bibfield  {author} {\bibinfo {author} {\bibfnamefont {S.}~\bibnamefont {Pancholi}},\ }\href@noop {} {\emph {\bibinfo {title} {Pear-shaped Nuclei}}}\ (\bibinfo  {publisher} {World Scientific Publishing Company},\ \bibinfo {year} {2020})\BibitemShut {NoStop}%
\bibitem [{\citenamefont {Nomura}\ \emph {et~al.}(2014)\citenamefont {Nomura}, \citenamefont {Vretenar}, \citenamefont {Nik\ifmmode \check{s}\else \v{s}\fi{}i\ifmmode~\acute{c}\else \'{c}\fi{}},\ and\ \citenamefont {Lu}}]{PhysRevC.89.024312}%
  \BibitemOpen
  \bibfield  {author} {\bibinfo {author} {\bibfnamefont {K.}~\bibnamefont {Nomura}}, \bibinfo {author} {\bibfnamefont {D.}~\bibnamefont {Vretenar}}, \bibinfo {author} {\bibfnamefont {T.}~\bibnamefont {Nik\ifmmode \check{s}\else \v{s}\fi{}i\ifmmode~\acute{c}\else \'{c}\fi{}}},\ and\ \bibinfo {author} {\bibfnamefont {B.-N.}\ \bibnamefont {Lu}},\ }\bibfield  {title} {\bibinfo {title} {Microscopic description of octupole shape-phase transitions in light actinide and rare-earth nuclei},\ }\href {https://doi.org/10.1103/PhysRevC.89.024312} {\bibfield  {journal} {\bibinfo  {journal} {Phys. Rev. C}\ }\textbf {\bibinfo {volume} {89}},\ \bibinfo {pages} {024312} (\bibinfo {year} {2014})}\BibitemShut {NoStop}%
\bibitem [{\citenamefont {Pospelov}\ and\ \citenamefont {Ritz}(2005)}]{POSPELOV2005119}%
  \BibitemOpen
  \bibfield  {author} {\bibinfo {author} {\bibfnamefont {M.}~\bibnamefont {Pospelov}}\ and\ \bibinfo {author} {\bibfnamefont {A.}~\bibnamefont {Ritz}},\ }\bibfield  {title} {\bibinfo {title} {Electric dipole moments as probes of new physics},\ }\href {https://doi.org/https://doi.org/10.1016/j.aop.2005.04.002} {\bibfield  {journal} {\bibinfo  {journal} {Annals of Physics}\ }\textbf {\bibinfo {volume} {318}},\ \bibinfo {pages} {119} (\bibinfo {year} {2005})},\ \bibinfo {note} {special Issue}\BibitemShut {NoStop}%
\bibitem [{\citenamefont {Griffith}\ \emph {et~al.}(2009)\citenamefont {Griffith}, \citenamefont {Swallows}, \citenamefont {Loftus}, \citenamefont {Romalis}, \citenamefont {Heckel},\ and\ \citenamefont {Fortson}}]{PhysRevLett.102.101601}%
  \BibitemOpen
  \bibfield  {author} {\bibinfo {author} {\bibfnamefont {W.~C.}\ \bibnamefont {Griffith}}, \bibinfo {author} {\bibfnamefont {M.~D.}\ \bibnamefont {Swallows}}, \bibinfo {author} {\bibfnamefont {T.~H.}\ \bibnamefont {Loftus}}, \bibinfo {author} {\bibfnamefont {M.~V.}\ \bibnamefont {Romalis}}, \bibinfo {author} {\bibfnamefont {B.~R.}\ \bibnamefont {Heckel}},\ and\ \bibinfo {author} {\bibfnamefont {E.~N.}\ \bibnamefont {Fortson}},\ }\bibfield  {title} {\bibinfo {title} {Improved limit on the permanent electric dipole moment of $^{199}\mathrm{Hg}$},\ }\href {https://doi.org/10.1103/PhysRevLett.102.101601} {\bibfield  {journal} {\bibinfo  {journal} {Phys. Rev. Lett.}\ }\textbf {\bibinfo {volume} {102}},\ \bibinfo {pages} {101601} (\bibinfo {year} {2009})}\BibitemShut {NoStop}%
\bibitem [{\citenamefont {Gaffney}\ \emph {et~al.}(2013)\citenamefont {Gaffney}, \citenamefont {Butler}, \citenamefont {Scheck}, \citenamefont {Hayes}, \citenamefont {Wenander}, \citenamefont {Albers}, \citenamefont {Bastin}, \citenamefont {Bauer}, \citenamefont {Blazhev}, \citenamefont {B{\"o}nig}, \citenamefont {Bree}, \citenamefont {Cederk{\"a}ll}, \citenamefont {Chupp}, \citenamefont {Cline}, \citenamefont {Cocolios}, \citenamefont {Davinson}, \citenamefont {De~Witte}, \citenamefont {Diriken}, \citenamefont {Grahn}, \citenamefont {Herzan}, \citenamefont {Huyse}, \citenamefont {Jenkins}, \citenamefont {Joss}, \citenamefont {Kesteloot}, \citenamefont {Konki}, \citenamefont {Kowalczyk}, \citenamefont {Kr{\"o}ll}, \citenamefont {Kwan}, \citenamefont {Lutter}, \citenamefont {Moschner}, \citenamefont {Napiorkowski}, \citenamefont {Pakarinen}, \citenamefont {Pfeiffer}, \citenamefont {Radeck}, \citenamefont {Reiter}, \citenamefont {Reynders}, \citenamefont {Rigby}, \citenamefont {Robledo}, \citenamefont
  {Rudigier}, \citenamefont {Sambi}, \citenamefont {Seidlitz}, \citenamefont {Siebeck}, \citenamefont {Stora}, \citenamefont {Thoele}, \citenamefont {Van~Duppen}, \citenamefont {Vermeulen}, \citenamefont {von Schmid}, \citenamefont {Voulot}, \citenamefont {Warr}, \citenamefont {Wimmer}, \citenamefont {Wrzosek-Lipska}, \citenamefont {Wu},\ and\ \citenamefont {Zielinska}}]{Gaffney2013-df}%
  \BibitemOpen
  \bibfield  {author} {\bibinfo {author} {\bibfnamefont {L.~P.}\ \bibnamefont {Gaffney}}, \bibinfo {author} {\bibfnamefont {P.~A.}\ \bibnamefont {Butler}}, \bibinfo {author} {\bibfnamefont {M.}~\bibnamefont {Scheck}}, \bibinfo {author} {\bibfnamefont {A.~B.}\ \bibnamefont {Hayes}}, \bibinfo {author} {\bibfnamefont {F.}~\bibnamefont {Wenander}}, \bibinfo {author} {\bibfnamefont {M.}~\bibnamefont {Albers}}, \bibinfo {author} {\bibfnamefont {B.}~\bibnamefont {Bastin}}, \bibinfo {author} {\bibfnamefont {C.}~\bibnamefont {Bauer}}, \bibinfo {author} {\bibfnamefont {A.}~\bibnamefont {Blazhev}}, \bibinfo {author} {\bibfnamefont {S.}~\bibnamefont {B{\"o}nig}}, \bibinfo {author} {\bibfnamefont {N.}~\bibnamefont {Bree}}, \bibinfo {author} {\bibfnamefont {J.}~\bibnamefont {Cederk{\"a}ll}}, \bibinfo {author} {\bibfnamefont {T.}~\bibnamefont {Chupp}}, \bibinfo {author} {\bibfnamefont {D.}~\bibnamefont {Cline}}, \bibinfo {author} {\bibfnamefont {T.~E.}\ \bibnamefont {Cocolios}}, \bibinfo {author} {\bibfnamefont
  {T.}~\bibnamefont {Davinson}}, \bibinfo {author} {\bibfnamefont {H.}~\bibnamefont {De~Witte}}, \bibinfo {author} {\bibfnamefont {J.}~\bibnamefont {Diriken}}, \bibinfo {author} {\bibfnamefont {T.}~\bibnamefont {Grahn}}, \bibinfo {author} {\bibfnamefont {A.}~\bibnamefont {Herzan}}, \bibinfo {author} {\bibfnamefont {M.}~\bibnamefont {Huyse}}, \bibinfo {author} {\bibfnamefont {D.~G.}\ \bibnamefont {Jenkins}}, \bibinfo {author} {\bibfnamefont {D.~T.}\ \bibnamefont {Joss}}, \bibinfo {author} {\bibfnamefont {N.}~\bibnamefont {Kesteloot}}, \bibinfo {author} {\bibfnamefont {J.}~\bibnamefont {Konki}}, \bibinfo {author} {\bibfnamefont {M.}~\bibnamefont {Kowalczyk}}, \bibinfo {author} {\bibfnamefont {T.}~\bibnamefont {Kr{\"o}ll}}, \bibinfo {author} {\bibfnamefont {E.}~\bibnamefont {Kwan}}, \bibinfo {author} {\bibfnamefont {R.}~\bibnamefont {Lutter}}, \bibinfo {author} {\bibfnamefont {K.}~\bibnamefont {Moschner}}, \bibinfo {author} {\bibfnamefont {P.}~\bibnamefont {Napiorkowski}}, \bibinfo {author} {\bibfnamefont
  {J.}~\bibnamefont {Pakarinen}}, \bibinfo {author} {\bibfnamefont {M.}~\bibnamefont {Pfeiffer}}, \bibinfo {author} {\bibfnamefont {D.}~\bibnamefont {Radeck}}, \bibinfo {author} {\bibfnamefont {P.}~\bibnamefont {Reiter}}, \bibinfo {author} {\bibfnamefont {K.}~\bibnamefont {Reynders}}, \bibinfo {author} {\bibfnamefont {S.~V.}\ \bibnamefont {Rigby}}, \bibinfo {author} {\bibfnamefont {L.~M.}\ \bibnamefont {Robledo}}, \bibinfo {author} {\bibfnamefont {M.}~\bibnamefont {Rudigier}}, \bibinfo {author} {\bibfnamefont {S.}~\bibnamefont {Sambi}}, \bibinfo {author} {\bibfnamefont {M.}~\bibnamefont {Seidlitz}}, \bibinfo {author} {\bibfnamefont {B.}~\bibnamefont {Siebeck}}, \bibinfo {author} {\bibfnamefont {T.}~\bibnamefont {Stora}}, \bibinfo {author} {\bibfnamefont {P.}~\bibnamefont {Thoele}}, \bibinfo {author} {\bibfnamefont {P.}~\bibnamefont {Van~Duppen}}, \bibinfo {author} {\bibfnamefont {M.~J.}\ \bibnamefont {Vermeulen}}, \bibinfo {author} {\bibfnamefont {M.}~\bibnamefont {von Schmid}}, \bibinfo {author}
  {\bibfnamefont {D.}~\bibnamefont {Voulot}}, \bibinfo {author} {\bibfnamefont {N.}~\bibnamefont {Warr}}, \bibinfo {author} {\bibfnamefont {K.}~\bibnamefont {Wimmer}}, \bibinfo {author} {\bibfnamefont {K.}~\bibnamefont {Wrzosek-Lipska}}, \bibinfo {author} {\bibfnamefont {C.~Y.}\ \bibnamefont {Wu}},\ and\ \bibinfo {author} {\bibfnamefont {M.}~\bibnamefont {Zielinska}},\ }\bibfield  {title} {\bibinfo {title} {Studies of pear-shaped nuclei using accelerated radioactive beams},\ }\href@noop {} {\bibfield  {journal} {\bibinfo  {journal} {Nature}\ }\textbf {\bibinfo {volume} {497}},\ \bibinfo {pages} {199} (\bibinfo {year} {2013})}\BibitemShut {NoStop}%
\bibitem [{\citenamefont {Thorsteinsen}\ \emph {et~al.}(1990)\citenamefont {Thorsteinsen}, \citenamefont {Nybø},\ and\ \citenamefont {Løvhøiden}}]{TFThorsteinsen_1990}%
  \BibitemOpen
  \bibfield  {author} {\bibinfo {author} {\bibfnamefont {T.~F.}\ \bibnamefont {Thorsteinsen}}, \bibinfo {author} {\bibfnamefont {K.}~\bibnamefont {Nybø}},\ and\ \bibinfo {author} {\bibfnamefont {G.}~\bibnamefont {Løvhøiden}},\ }\bibfield  {title} {\bibinfo {title} {Levels in 226ra populated by inelastic deuteron scattering},\ }\href {https://doi.org/10.1088/0031-8949/42/2/004} {\bibfield  {journal} {\bibinfo  {journal} {Physica Scripta}\ }\textbf {\bibinfo {volume} {42}},\ \bibinfo {pages} {141} (\bibinfo {year} {1990})}\BibitemShut {NoStop}%
\bibitem [{\citenamefont {Cseh}(2015)}]{CSEH2015213}%
  \BibitemOpen
  \bibfield  {author} {\bibinfo {author} {\bibfnamefont {J.}~\bibnamefont {Cseh}},\ }\bibfield  {title} {\bibinfo {title} {Algebraic models for shell-like quarteting of nucleons},\ }\href {https://doi.org/https://doi.org/10.1016/j.physletb.2015.02.034} {\bibfield  {journal} {\bibinfo  {journal} {Physics Letters B}\ }\textbf {\bibinfo {volume} {743}},\ \bibinfo {pages} {213} (\bibinfo {year} {2015})}\BibitemShut {NoStop}%
\bibitem [{\citenamefont {Bonatsos}\ \emph {et~al.}(2017{\natexlab{b}})\citenamefont {Bonatsos}, \citenamefont {Assimakis}, \citenamefont {Minkov}, \citenamefont {Martinou}, \citenamefont {Cakirli}, \citenamefont {Casten},\ and\ \citenamefont {Blaum}}]{PhysRevC.95.064325}%
  \BibitemOpen
  \bibfield  {author} {\bibinfo {author} {\bibfnamefont {D.}~\bibnamefont {Bonatsos}}, \bibinfo {author} {\bibfnamefont {I.~E.}\ \bibnamefont {Assimakis}}, \bibinfo {author} {\bibfnamefont {N.}~\bibnamefont {Minkov}}, \bibinfo {author} {\bibfnamefont {A.}~\bibnamefont {Martinou}}, \bibinfo {author} {\bibfnamefont {R.~B.}\ \bibnamefont {Cakirli}}, \bibinfo {author} {\bibfnamefont {R.~F.}\ \bibnamefont {Casten}},\ and\ \bibinfo {author} {\bibfnamefont {K.}~\bibnamefont {Blaum}},\ }\bibfield  {title} {\bibinfo {title} {Proxy-su(3) symmetry in heavy deformed nuclei},\ }\href {https://doi.org/10.1103/PhysRevC.95.064325} {\bibfield  {journal} {\bibinfo  {journal} {Phys. Rev. C}\ }\textbf {\bibinfo {volume} {95}},\ \bibinfo {pages} {064325} (\bibinfo {year} {2017}{\natexlab{b}})}\BibitemShut {NoStop}%
\bibitem [{\citenamefont {Bonatsos}\ \emph {et~al.}(2023)\citenamefont {Bonatsos}, \citenamefont {Martinou}, \citenamefont {Peroulis}, \citenamefont {Mertzimekis},\ and\ \citenamefont {Minkov}}]{sym15010169}%
  \BibitemOpen
  \bibfield  {author} {\bibinfo {author} {\bibfnamefont {D.}~\bibnamefont {Bonatsos}}, \bibinfo {author} {\bibfnamefont {A.}~\bibnamefont {Martinou}}, \bibinfo {author} {\bibfnamefont {S.~K.}\ \bibnamefont {Peroulis}}, \bibinfo {author} {\bibfnamefont {T.~J.}\ \bibnamefont {Mertzimekis}},\ and\ \bibinfo {author} {\bibfnamefont {N.}~\bibnamefont {Minkov}},\ }\bibfield  {title} {\bibinfo {title} {The proxy-su(3) symmetry in atomic nuclei},\ }\bibfield  {journal} {\bibinfo  {journal} {Symmetry}\ }\textbf {\bibinfo {volume} {15}},\ \href {https://doi.org/10.3390/sym15010169} {10.3390/sym15010169} (\bibinfo {year} {2023})\BibitemShut {NoStop}%
\bibitem [{\citenamefont {Cseh}(2020)}]{PhysRevC.101.054306}%
  \BibitemOpen
  \bibfield  {author} {\bibinfo {author} {\bibfnamefont {J.}~\bibnamefont {Cseh}},\ }\bibfield  {title} {\bibinfo {title} {Shell-like quarteting in heavy nuclei: Algebraic approaches based on the pseudo- and proxy-su(3) schemes},\ }\href {https://doi.org/10.1103/PhysRevC.101.054306} {\bibfield  {journal} {\bibinfo  {journal} {Phys. Rev. C}\ }\textbf {\bibinfo {volume} {101}},\ \bibinfo {pages} {054306} (\bibinfo {year} {2020})}\BibitemShut {NoStop}%
\bibitem [{\citenamefont {Restrepo}\ and\ \citenamefont {Valencia}(2024)}]{restrepo2024}%
  \BibitemOpen
  \bibfield  {author} {\bibinfo {author} {\bibfnamefont {A.}~\bibnamefont {Restrepo}}\ and\ \bibinfo {author} {\bibfnamefont {J.~P.}\ \bibnamefont {Valencia}},\ }\href {https://arxiv.org/abs/2405.04679} {\bibinfo {title} {Inter-band b(e1) strengths in heavy nuclei based on the proxy-su(3) scheme}} (\bibinfo {year} {2024}),\ \Eprint {https://arxiv.org/abs/2405.04679} {arXiv:2405.04679 [nucl-th]} \BibitemShut {NoStop}%
\bibitem [{\citenamefont {Singh}\ and\ \citenamefont {Singh}()}]{nndc}%
  \BibitemOpen
  \bibfield  {author} {\bibinfo {author} {\bibfnamefont {B.}~\bibnamefont {Singh}}\ and\ \bibinfo {author} {\bibfnamefont {S.}~\bibnamefont {Singh}},\ }\href@noop {} {}\bibinfo {note} {From ENSDF database as of March 08, 2022. Version available at \url{http://www.nndc.bnl.gov/ensarchivals/}}\BibitemShut {NoStop}%
\bibitem [{\citenamefont {Martinou}\ and\ \citenamefont {Minkov}(2024)}]{Martinou_2024}%
  \BibitemOpen
  \bibfield  {author} {\bibinfo {author} {\bibfnamefont {A.}~\bibnamefont {Martinou}}\ and\ \bibinfo {author} {\bibfnamefont {N.}~\bibnamefont {Minkov}},\ }\bibfield  {title} {\bibinfo {title} {Microscopic derivation of the octupole magic numbers from symmetry considerations},\ }\href {https://doi.org/10.1088/1402-4896/ad562f} {\bibfield  {journal} {\bibinfo  {journal} {Physica Scripta}\ }\textbf {\bibinfo {volume} {99}},\ \bibinfo {pages} {075311} (\bibinfo {year} {2024})}\BibitemShut {NoStop}%
\bibitem [{\citenamefont {Isacker}\ and\ \citenamefont {Pittel}(2016)}]{Isacker_2016}%
  \BibitemOpen
  \bibfield  {author} {\bibinfo {author} {\bibfnamefont {P.~V.}\ \bibnamefont {Isacker}}\ and\ \bibinfo {author} {\bibfnamefont {S.}~\bibnamefont {Pittel}},\ }\bibfield  {title} {\bibinfo {title} {Symmetries and deformations in the spherical shell model},\ }\href {https://doi.org/10.1088/0031-8949/91/2/023009} {\bibfield  {journal} {\bibinfo  {journal} {Physica Scripta}\ }\textbf {\bibinfo {volume} {91}},\ \bibinfo {pages} {023009} (\bibinfo {year} {2016})}\BibitemShut {NoStop}%
\bibitem [{\citenamefont {Frank}\ \emph {et~al.}(2008)\citenamefont {Frank}, \citenamefont {Jolie},\ and\ \citenamefont {van Isacker}}]{frank2008symmetries}%
  \BibitemOpen
  \bibfield  {author} {\bibinfo {author} {\bibfnamefont {A.}~\bibnamefont {Frank}}, \bibinfo {author} {\bibfnamefont {J.}~\bibnamefont {Jolie}},\ and\ \bibinfo {author} {\bibfnamefont {P.}~\bibnamefont {van Isacker}},\ }\href@noop {} {\emph {\bibinfo {title} {Symmetries in Atomic Nuclei: From Isospin to Supersymmetry}}},\ Springer Tracts in Modern Physics\ (\bibinfo  {publisher} {Springer New York},\ \bibinfo {year} {2008})\BibitemShut {NoStop}%
\bibitem [{\citenamefont {Draayer}\ \emph {et~al.}(1989)\citenamefont {Draayer}, \citenamefont {Leschber}, \citenamefont {Park},\ and\ \citenamefont {Lopez}}]{DRAAYER1989279}%
  \BibitemOpen
  \bibfield  {author} {\bibinfo {author} {\bibfnamefont {J.}~\bibnamefont {Draayer}}, \bibinfo {author} {\bibfnamefont {Y.}~\bibnamefont {Leschber}}, \bibinfo {author} {\bibfnamefont {S.}~\bibnamefont {Park}},\ and\ \bibinfo {author} {\bibfnamefont {R.}~\bibnamefont {Lopez}},\ }\bibfield  {title} {\bibinfo {title} {Representations of u(3) in u(n)},\ }\href {https://doi.org/https://doi.org/10.1016/0010-4655(89)90024-6} {\bibfield  {journal} {\bibinfo  {journal} {Computer Physics Communications}\ }\textbf {\bibinfo {volume} {56}},\ \bibinfo {pages} {279} (\bibinfo {year} {1989})}\BibitemShut {NoStop}%
\bibitem [{\citenamefont {Sloane}(2014)}]{sloane2014lie}%
  \BibitemOpen
  \bibfield  {author} {\bibinfo {author} {\bibfnamefont {F.}~\bibnamefont {Sloane}},\ }\href@noop {} {\emph {\bibinfo {title} {Lie Algebras and Applications}}},\ Lecture Notes in Physics\ (\bibinfo  {publisher} {Springer Berlin Heidelberg},\ \bibinfo {year} {2014})\BibitemShut {NoStop}%
\bibitem [{\citenamefont {Vergados}(1968)}]{VERGADOS1968681}%
  \BibitemOpen
  \bibfield  {author} {\bibinfo {author} {\bibfnamefont {J.}~\bibnamefont {Vergados}},\ }\bibfield  {title} {\bibinfo {title} {Su(3) r(3) wigner coefficients in the 2s-1d shell},\ }\href {https://doi.org/https://doi.org/10.1016/0375-9474(68)90249-2} {\bibfield  {journal} {\bibinfo  {journal} {Nuclear Physics A}\ }\textbf {\bibinfo {volume} {111}},\ \bibinfo {pages} {681} (\bibinfo {year} {1968})}\BibitemShut {NoStop}%
\bibitem [{\citenamefont {Moshinsky}\ and\ \citenamefont {Smirnov}(1996)}]{moshinsky1996harmonic}%
  \BibitemOpen
  \bibfield  {author} {\bibinfo {author} {\bibfnamefont {M.}~\bibnamefont {Moshinsky}}\ and\ \bibinfo {author} {\bibfnamefont {Y.}~\bibnamefont {Smirnov}},\ }\href@noop {} {\emph {\bibinfo {title} {The Harmonic Oscillator in Modern Physics}}},\ Contemporary concepts in physics\ (\bibinfo  {publisher} {Harwood Academic Publishers},\ \bibinfo {year} {1996})\BibitemShut {NoStop}%
\bibitem [{\citenamefont {Troltenier}\ \emph {et~al.}(1995)\citenamefont {Troltenier}, \citenamefont {Bahri},\ and\ \citenamefont {Draayer}}]{TROLTENIER199553}%
  \BibitemOpen
  \bibfield  {author} {\bibinfo {author} {\bibfnamefont {D.}~\bibnamefont {Troltenier}}, \bibinfo {author} {\bibfnamefont {C.}~\bibnamefont {Bahri}},\ and\ \bibinfo {author} {\bibfnamefont {J.}~\bibnamefont {Draayer}},\ }\bibfield  {title} {\bibinfo {title} {Generalized pseudo-su(3) model and pairing},\ }\href {https://doi.org/https://doi.org/10.1016/0375-9474(94)00518-R} {\bibfield  {journal} {\bibinfo  {journal} {Nuclear Physics A}\ }\textbf {\bibinfo {volume} {586}},\ \bibinfo {pages} {53} (\bibinfo {year} {1995})}\BibitemShut {NoStop}%
\bibitem [{\citenamefont {Kota}(2020)}]{kota20203}%
  \BibitemOpen
  \bibfield  {author} {\bibinfo {author} {\bibfnamefont {V.}~\bibnamefont {Kota}},\ }\href@noop {} {\emph {\bibinfo {title} {SU(3) Symmetry in Atomic Nuclei}}}\ (\bibinfo  {publisher} {Springer Nature Singapore},\ \bibinfo {year} {2020})\BibitemShut {NoStop}%
\bibitem [{\citenamefont {Brussaard}\ and\ \citenamefont {Glaudemans}(1977)}]{brussaard1977shell}%
  \BibitemOpen
  \bibfield  {author} {\bibinfo {author} {\bibfnamefont {P.}~\bibnamefont {Brussaard}}\ and\ \bibinfo {author} {\bibfnamefont {P.}~\bibnamefont {Glaudemans}},\ }\href@noop {} {\emph {\bibinfo {title} {Shell-model Applications in Nuclear Spectroscopy}}}\ (\bibinfo  {publisher} {North-Holland Publishing Company},\ \bibinfo {year} {1977})\BibitemShut {NoStop}%
\bibitem [{\citenamefont {Leander}\ and\ \citenamefont {Sheline}(1984)}]{LEANDER1984375}%
  \BibitemOpen
  \bibfield  {author} {\bibinfo {author} {\bibfnamefont {G.}~\bibnamefont {Leander}}\ and\ \bibinfo {author} {\bibfnamefont {R.}~\bibnamefont {Sheline}},\ }\bibfield  {title} {\bibinfo {title} {Intrinsic reflection asymmetry in odd-a nuclei},\ }\href {https://doi.org/https://doi.org/10.1016/0375-9474(84)90417-2} {\bibfield  {journal} {\bibinfo  {journal} {Nuclear Physics A}\ }\textbf {\bibinfo {volume} {413}},\ \bibinfo {pages} {375} (\bibinfo {year} {1984})}\BibitemShut {NoStop}%
\bibitem [{\citenamefont {Möller}\ \emph {et~al.}(2016)\citenamefont {Möller}, \citenamefont {Sierk}, \citenamefont {Ichikawa},\ and\ \citenamefont {Sagawa}}]{MOLLER20161}%
  \BibitemOpen
  \bibfield  {author} {\bibinfo {author} {\bibfnamefont {P.}~\bibnamefont {Möller}}, \bibinfo {author} {\bibfnamefont {A.}~\bibnamefont {Sierk}}, \bibinfo {author} {\bibfnamefont {T.}~\bibnamefont {Ichikawa}},\ and\ \bibinfo {author} {\bibfnamefont {H.}~\bibnamefont {Sagawa}},\ }\bibfield  {title} {\bibinfo {title} {Nuclear ground-state masses and deformations: Frdm(2012)},\ }\href {https://doi.org/https://doi.org/10.1016/j.adt.2015.10.002} {\bibfield  {journal} {\bibinfo  {journal} {Atomic Data and Nuclear Data Tables}\ }\textbf {\bibinfo {volume} {109-110}},\ \bibinfo {pages} {1} (\bibinfo {year} {2016})}\BibitemShut {NoStop}%
\bibitem [{\citenamefont {Robledo}\ and\ \citenamefont {Butler}(2013)}]{PhysRevC.88.051302}%
  \BibitemOpen
  \bibfield  {author} {\bibinfo {author} {\bibfnamefont {L.~M.}\ \bibnamefont {Robledo}}\ and\ \bibinfo {author} {\bibfnamefont {P.~A.}\ \bibnamefont {Butler}},\ }\bibfield  {title} {\bibinfo {title} {Quadrupole-octupole coupling in the light actinides},\ }\href {https://doi.org/10.1103/PhysRevC.88.051302} {\bibfield  {journal} {\bibinfo  {journal} {Phys. Rev. C}\ }\textbf {\bibinfo {volume} {88}},\ \bibinfo {pages} {051302} (\bibinfo {year} {2013})}\BibitemShut {NoStop}%
\bibitem [{\citenamefont {Gaffney}(2012)}]{gaffney}%
  \BibitemOpen
  \bibfield  {author} {\bibinfo {author} {\bibfnamefont {L.~P.}\ \bibnamefont {Gaffney}},\ }\emph {\bibinfo {title} {Octupole collectivity in 220Rn and 224Ra}},\ \href@noop {} {\bibinfo {type} {Phd thesis}},\ \bibinfo  {school} {University of Liverpool} (\bibinfo {year} {2012}),\ \bibinfo {note} {available at \url{https://cds.cern.ch/record/1547569/files/CERN-THESIS-2012-309.pdf}}\BibitemShut {NoStop}%
\bibitem [{\citenamefont {Wybourne}(2022)}]{schurprogram}%
  \BibitemOpen
  \bibfield  {author} {\bibinfo {author} {\bibfnamefont {B.~G.}\ \bibnamefont {Wybourne}},\ }\href {https://schur.sourceforge.net/} {\bibinfo {title} {{Schur Group Theory Software}}} (\bibinfo {year} {2022})\BibitemShut {NoStop}%
\bibitem [{\citenamefont {Dytrych}\ \emph {et~al.}(2021)\citenamefont {Dytrych}, \citenamefont {Langr}, \citenamefont {Draayer}, \citenamefont {Launey},\ and\ \citenamefont {Gazda}}]{DYTRYCH2021108137}%
  \BibitemOpen
  \bibfield  {author} {\bibinfo {author} {\bibfnamefont {T.}~\bibnamefont {Dytrych}}, \bibinfo {author} {\bibfnamefont {D.}~\bibnamefont {Langr}}, \bibinfo {author} {\bibfnamefont {J.~P.}\ \bibnamefont {Draayer}}, \bibinfo {author} {\bibfnamefont {K.~D.}\ \bibnamefont {Launey}},\ and\ \bibinfo {author} {\bibfnamefont {D.}~\bibnamefont {Gazda}},\ }\bibfield  {title} {\bibinfo {title} {Su3lib: A c++ library for accurate computation of wigner and racah coefficients of su(3)},\ }\href {https://doi.org/https://doi.org/10.1016/j.cpc.2021.108137} {\bibfield  {journal} {\bibinfo  {journal} {Computer Physics Communications}\ }\textbf {\bibinfo {volume} {269}},\ \bibinfo {pages} {108137} (\bibinfo {year} {2021})}\BibitemShut {NoStop}%
\bibitem [{\citenamefont {Bahri}\ and\ \citenamefont {Draayer}(1994)}]{BAHRI199459}%
  \BibitemOpen
  \bibfield  {author} {\bibinfo {author} {\bibfnamefont {C.}~\bibnamefont {Bahri}}\ and\ \bibinfo {author} {\bibfnamefont {J.}~\bibnamefont {Draayer}},\ }\bibfield  {title} {\bibinfo {title} {Su(3) reduced matrix element package},\ }\href {https://doi.org/https://doi.org/10.1016/0010-4655(94)90035-3} {\bibfield  {journal} {\bibinfo  {journal} {Computer Physics Communications}\ }\textbf {\bibinfo {volume} {83}},\ \bibinfo {pages} {59} (\bibinfo {year} {1994})}\BibitemShut {NoStop}%
\bibitem [{\citenamefont {Millener}(1991)}]{Millener1991SU3IS}%
  \BibitemOpen
  \bibfield  {author} {\bibinfo {author} {\bibfnamefont {D.~J.}\ \bibnamefont {Millener}},\ }\bibfield  {title} {\bibinfo {title} {Su(3) in shell-model calculations}\ }(\bibinfo {year} {1991})\BibitemShut {NoStop}%
\bibitem [{\citenamefont {Hirsch}\ \emph {et~al.}(2007)\citenamefont {Hirsch}, \citenamefont {Bahri}, \citenamefont {Draayer}, \citenamefont {Castaños},\ and\ \citenamefont {Hess}}]{J2007ReducedME}%
  \BibitemOpen
  \bibfield  {author} {\bibinfo {author} {\bibfnamefont {G.~J.}\ \bibnamefont {Hirsch}}, \bibinfo {author} {\bibfnamefont {C.}~\bibnamefont {Bahri}}, \bibinfo {author} {\bibfnamefont {J.}~\bibnamefont {Draayer}}, \bibinfo {author} {\bibfnamefont {O.}~\bibnamefont {Castaños}},\ and\ \bibinfo {author} {\bibfnamefont {P.}~\bibnamefont {Hess}},\ }\bibfield  {title} {\bibinfo {title} {Reduced matrix elements for the leading spin zero states in the su ( 3 ) scheme *}\ }(\bibinfo {year} {2007})\BibitemShut {NoStop}%
\bibitem [{\citenamefont {Suhonen}(2007)}]{suhonen2007nucleons}%
  \BibitemOpen
  \bibfield  {author} {\bibinfo {author} {\bibfnamefont {J.}~\bibnamefont {Suhonen}},\ }\href@noop {} {\emph {\bibinfo {title} {From Nucleons to Nucleus: Concepts of Microscopic Nuclear Theory}}},\ Theoretical and Mathematical Physics\ (\bibinfo  {publisher} {Springer Berlin Heidelberg},\ \bibinfo {year} {2007})\BibitemShut {NoStop}%
\end{thebibliography}%

\end{document}